\documentclass[aps,prd,final,twocolumn,letterpaper]{revtex4}

\usepackage{graphicx}                            
\usepackage{epstopdf}
\usepackage{amsmath} 
\usepackage{amssymb}
\usepackage{float}
\usepackage{amsthm}
\usepackage{enumerate}
\usepackage{subfig}
\usepackage[all]{nowidow}

\usepackage[usenames,dvipsnames]{color}

\newcommand{\bra}[1]{\left\langle #1 \right|}
\newcommand{\ket}[1]{\left|#1\right\rangle}
\newcommand{\braket}[2]{\left\langle#1 |  #2\right\rangle}

\newcommand{\bigoh}[1]{\mathcal{O} (#1)}
\newcounter{theoremcount}
\newcounter{lemmacount}
\newenvironment{maintheorem}
	{\refstepcounter{theoremcount}
    \vspace{12 pt}
    \noindent
	\textbf{Theorem \thetheoremcount}.
    }{
    \vspace{12pt}
    }
\newenvironment{lemma}
	{\refstepcounter{lemmacount}
    \vspace{12 pt}
    \noindent
	\textbf{Lemma \thelemmacount}.
    }{
    \vspace{12pt}
    }
\newenvironment{mainproof}[1]
	{\vspace{12 pt}
    \noindent
	\textbf{#1}.
    }{\hfill \QEDA
    \vspace{12pt}
    }
\newenvironment{definition}[1]
	{\vspace{12 pt}
    \noindent
	\textbf{#1}.
    }{
    \vspace{12pt}
    }
    
\newcommand*{\QEDA}{\hfill\ensuremath{\blacksquare}}%

\begin{document}

\title{Fixed-Point Adiabatic Quantum Search}
\author{Alexander~M.~Dalzell}
\author{Theodore~J.~Yoder}
\author{Isaac~L.~Chuang}
\affiliation{Massachusetts Institute of Technology, Cambridge, Massachusetts 02139, USA}
\date{\today} 

\begin{abstract}

\noindent Fixed-point quantum search algorithms succeed at finding one of $M$ target items among $N$ total items even when the run time of the algorithm is longer than necessary. While the famous Grover's algorithm can search quadratically faster than a classical computer, it lacks the fixed-point property --- the fraction of target items must be known precisely to know when to terminate the algorithm. Recently, Yoder et~al.~\cite{yoder2014fixed} gave an optimal gate-model search algorithm with the fixed-point property. Meanwhile, it is known \cite{roland2002quantum} that an \emph{adiabatic} quantum algorithm, operating by continuously varying a Hamiltonian, can reproduce the quadratic speedup of gate-model Grover search. We ask, can an adiabatic algorithm also reproduce the fixed-point property? We show that the answer depends on what interpolation schedule is used, so as in the gate model, there are both fixed-point and non-fixed-point versions of adiabatic search, only some of which attain the quadratic quantum speedup. Guided by geometric intuition on the Bloch sphere, we rigorously justify our claims with an explicit upper bound on the error in the adiabatic approximation.
We also show that the fixed-point adiabatic search algorithm can be simulated in the gate model with neither loss of the quadratic Grover speedup nor of the fixed-point property. Finally, we discuss natural uses of fixed-point algorithms such as preparation of a relatively prime state and oblivious amplitude amplification.

\end{abstract}

\maketitle

\pagestyle{myheadings}
\markboth{A.M. Dalzell}{Fixed-Point Adiabatic Quantum Search}
\thispagestyle{empty}


\section{Introduction}\label{sec:introduction}
The remarkable discovery of Grover's algorithm \cite{grover1996fast}, which solves the simplest of search problems quadratically faster than a classical computer, has helped fuel interest in quantum computing for the last two decades. 

Grover's algorithm finds one of $M$ target items among $N$ total items in $\bigoh{\sqrt{N/M}}$ time with high probability by repeated application of an operation called the Grover iterate on an initial superposition of all of the items. The quantum state lies in an $N$ dimensional Hilbert space, but the symmetry of the problem allows for a reduction to two dimensions. In this picture, the Grover iterate can be interpreted as a rotation which moves the quantum state closer to the direction of the target items. However, too many applications causes the state to overrotate; knowledge of $\lambda=M/N$, the fraction of target items, is needed to choose the right number of iterations. 

How do we find a target item when $\lambda$ is unknown, while maintaining the Grover-like quadratic speedup? One solution is to first estimate $\lambda$ using quantum counting algorithms \cite{brassard1998quantum,brassard2002quantum,hen2014fast}; another involves choosing the number of iterations randomly from an exponentially increasing range of integers until a target state is found \cite{boyer1998tight}. Although these methods have Grover-like scaling, they require measuring the system and repreparing the initial state. In contrast, Aaronson and Christiano \cite{aaronson2012quantum} give a solution which prepares a state, albeit a mixed state, arbitrarily near to the target state subspace avoiding the need for measurement and requiring only knowledge of a (typically small) lower bound $w$ for the fraction $\lambda$. Similarly, under the same assumption $w \leq \lambda$, Yoder~et~al. \cite{yoder2014fixed} provide an algorithm which coherently prepares the (pure) uniform superposition $\ket E$ over the $M$ target states with arbitrarily small error: it produces $\ket E$ with fidelity at least $\sqrt{1-\delta^2}$ in only $\bigoh{\log(1/\delta)/\sqrt{w}}$ time, thus exhibiting the quadratic speedup. The algorithm is a \textit{fixed-point} search algorithm \cite{grover2005fixed, tulsi2006new} since the success probability remains high even when $\lambda > w$ (i.e.~the algorithm was run longer than necessary).

Fixed-point search algorithms can be used for error-correcting schemes \cite{reichardt2005quantum}, oblivious amplitude amplification \cite{berry2014exponential}, or situations where a natural lower bound on $\lambda$ exists, such as preparation of the relatively-prime state (see section \ref{sec:applications}). More generally, knowing the existence of even one target state provides a lower bound $\lambda \geq 1/N$. Additionally, Aaronson and Christiano \cite{aaronson2012quantum} explain how a fixed-point algorithm might be a strategy for counterfeiters of quantum money to amplify imperfect copies. 


The $\bigoh{\sqrt{N/M}}$ scaling of Grover's algorithm has been shown to be optimal \cite{bennett1997strengths}, but this speedup is not restricted to the gate model; a similar speedup is found in continuous models. For example, Farhi and Gutmann introduced the Hamiltonian oracle model \cite{farhi1998analog} and showed how an analogue of search displays the quadratic speedup.
Likewise, the speedup is reproduced \cite{roland2002quantum} for searching in the model of Adiabatic Quantum Computation (AQC) \cite{farhi2000quantum}. But these algorithms rely on knowledge of $\lambda$, the first to decide when to terminate the evolution and the second for defining the adiabatic interpolation schedule. Since Hamiltonians generate rotations, we might expect to be hindered by the same problem of overrotation that plagues Grover's algorithm in the absence of knowledge of $\lambda$. Indeed, search in the Hamiltonian oracle model faces this exact problem. However, for AQC, we will show how this issue is avoided by presenting a fixed-point version of the adiabatic search algorithm with run time which displays Grover-like scaling (meanwhile, other versions are shown not to share these properties). 


In the context of AQC, fixed-point algorithms imply a robustness to systematic errors in the initial Hamiltonian. Moreover, a continuous algorithm might fit naturally into experiment, or perhaps provide a constant-factor speedup in implementation over a gate-model fixed-point algorithm. Indeed, implementing fixed-point search as a continuous algorithm, we gain continuous control over the lower bound $w$, which in the gate-model algorithm \cite{yoder2014fixed} is related to the number of Grover iterates and is therefore discrete.

Rigorously proving that a version of the adiabatic search algorithm is fixed point requires precise statements about the adiabatic theorem, which are often elusive. Many AQC works simply take a heuristic approach to the adiabatic theorem \cite{farhi2000quantum,farhi2001quantum}, and do not explicitly evaluate or bound the error in the adiabatic approximation, instead declaring the approximation valid if the Hamiltonian varies slowly in comparison to the square of its eigenvalue gap. Other works, including \cite{roland2002quantum}, use explicit bounds on the error which have since been shown to be incorrect in some cases \cite{marzlin2004inconsistency,tong2005quantitative}. More rigorous approaches have been taken. For example, in \cite{avron1987adiabatic,hagedorn2002elementary,joye1991full,jansen2007bounds,lidar2009adiabatic,rezakhani2010accuracy}, the error is written as a series in powers of $1/T$ where $T$ is the total run time. Rezakhani~et~al. \cite{rezakhani2010accuracy} do this explicitly for the adiabatic search algorithm, while some of the other treatments rely on assumptions or consider cases which are not applicable to the adiabatic search algorithm. This type of expansion is useful for understanding the asymptotic behavior of the algorithm, but less so for finding explicit bounds on the error when $T$ and $\lambda$ are finite. In \cite{reichardt2004quantum} and \cite{ambainis2004elementary}, a bound on the error is proved in terms of the eigenvalue gaps and derivatives of the Hamiltonian, but these works are concerned mostly with establishing a polynomial relationship between $T$, the gap, and the adiabatic error. These bounds are not tight enough to imply Grover-like scaling for the search algorithm. It would be preferable to develop an intuitively-motivated and explicit upper bound on the algorithm's failure probability which is both tight enough to give reliable scaling estimates and easy to evaluate for finite run times of the adiabatic search algorithm.

We develop exactly such a method for analysis of the adiabatic search algorithm motivated by the geometric intuition that a two-level system evolving by a time-dependent Hamiltonian is equivalent to a spin precessing about a varying magnetic field. Using this framework we derive three main results. 

First, we consider an arbitrary interpolation schedule and find an explicit expression (Theorem \ref{thm:generalbound}) which upper bounds the failure probability of the algorithm. We evaluate the upper bound for several different families of schedules, parameterized by $w$, a lower bound for $\lambda$, and $\epsilon$, a slowness parameter. We show examples of schedules which have the fixed-point property but lack Grover-like scaling, and vice-versa. The ``standard" schedule considered by \cite{roland2002quantum} is shown to have both Grover-like scaling and the fixed-point property simultaneously (Theorem \ref{thm:standardfixedpoint}).

Second, we modify our argument to produce an exact expression for the failure probability for a particular schedule (Theorem \ref{thm:fastexact}), which reproduces the quantum speedup for search -- in fact it is even faster than the standard schedule by a constant factor -- but is not fixed point. Exact expressions such as this are a rare occurrence among results regarding the adiabatic theorem.

Third, we describe a gate-model algorithm in the same oracle framework of \cite{yoder2014fixed} which simulates the adiabatic search algorithm and we bound its failure probability (Theorem \ref{thm:simulatedgeneralbound}). This bound implies that the simulation of the adiabatic search algorithm with standard schedule retains the fixed-point property and Grover-like scaling, yielding an alternative gate-model fixed-point algorithm. We argue that the existence of such a gate-model simulation that does not compromise the quantum speedup is not obvious.

\section{The Adiabatic Search Algorithm}\label{sec:adiabaticsearchalgorithm}
In this section, we set up the adiabatic search problem and discuss the adiabatic theorem. Continuous-time search takes place in an $N$-dimensional Hilbert space, a superposition of the orthonormal states $\ket 0, \ket 1, \ldots \ket {N-1}$. Some number, $M$, of these states have been marked, dubbed \emph{target states}, and our goal is to obtain one of them.
To do so, we are given access to (scaled multiples of) the oracle Hamiltonian $I-P$ where $P$ is the projector onto the space spanned by the $M$ target states, and $I$ is the identity operator. We see that any target state is an eigenstate of $I-P$ with eigenvalue 0, while any non-target state is an eigenstate with eigenvalue 1. This is not the only oracle Hamiltonian which works; any Hamiltonian whose ground state is a solution to the search problem is sufficient.

The adiabatic solution to this search problem begins the system in the equal-superposition state
\begin{equation}
\ket B = \frac{1}{\sqrt{N}}\sum_{x=0}^{N-1} \ket x
\label{eq:equalsuperposition}
\end{equation}
at time $t=0$, and further applies the Hamiltonian
\begin{equation}
H(t) = (1-s(t))H_0 + s(t)H_1
\label{eq:H}
\end{equation}
where 
\begin{eqnarray}
H_0 &=& I-\ket B \bra B \label{eq:H0} \\
H_1 &=& I-P \label{eq:H1}
\end{eqnarray}
and $s(t)$ is a continuous and monotonically nondecreasing function of time such that $s(0)=0$ and $s(T)=1$. We see that the beginning state $\ket B$ is the ground state of the initial Hamiltonian $H_0$. We are assuming, as is standard \cite{farhi1998analog,farhi2000quantum}, that the ability to apply certain Hamiltonians individually also implies the ability to apply their sum. Experimentally this is quite justifiable (see e.g.~\cite{steffen2003experimental,dickson2013thermally}).

With this physical setup, the system, represented by $\ket {\psi(t)}$ evolves based on the Schr{\"o}dinger equation, which is given by
\begin{equation}
i \frac{d}{dt}\ket{\psi(t)} = H(t) \ket {\psi(t)}.
\label{eq:schrodinger}
\end{equation}
The adiabatic theorem suggests that if the Hamiltonian is changed slowly enough, the system will remain approximately in the instantaneous ground state of the Hamiltonian. If this is true, at time $t=T$, the system will be approximately in the ground state of $H_1$, and when measured will yield a target state with high probability. 
%
%
\subsection{The Adiabatic Theorem}\label{sec:adiabatictheorem}
The adiabatic theorem is used to approximate the evolution of a quantum state subject to a continuously varying Hamiltonian. Suppose a system represented by $\ket{\psi(t)}$ is exposed to a time-dependent Hamiltonian $H(t)$, as is true for the algorithm we discuss in this paper. Further, suppose $H(t)$ has spectral decomposition
\begin{equation}
H(t)\ket {\phi_n(t)} = E_n(t) \ket {\phi_n(t)}
\label{eq:spectraldecomposition}
\end{equation}
where $E_n(t)$ and $\ket {\phi_n(t)}$ are the instantaneous eigenenergies and eigenstates of the Hamiltonian, with $E_0 < E_1 \ldots < E_N$. 
The adiabatic theorem states that if the state begins in the ground state $\ket {\psi(0)} = \ket {\phi_0(0)}$, then it will stay approximately in the ground state (up to a global phase) throughout the evolution $\ket {\psi(t)} \approx \ket {\phi_0(t)}$, provided that the Hamiltonian is changing sufficiently slowly. The criteria for slowness has been a notorious point of confusion for AQC. The general heuristic for adiabaticity used in early papers on AQC, such as \cite{farhi2000quantum}, relates the rate of change of the Hamiltonian to the energy gap $\Delta(t) \equiv E_1(t)-E_0(t)$ between the ground state and the first excited state
\begin{equation}
\lvert \bra {\phi_1(t)} \dot H(t) \ket {\phi_0(t)} \rvert \ll \Delta(t)^2.
\label{eq:adiabaticcondition}
\end{equation}
This makes the notion of slowness less vague, but still tells us nothing about how exactly the slowness of the algorithm affects the error in the approximation. Moreover, a closer examination reveals this criteria is not even completely correct \cite{marzlin2004inconsistency,tong2005quantitative}. However, more rigorous treatments of the adiabatic theorem have been done \cite{avron1987adiabatic,hagedorn2002elementary,joye1991full,reichardt2004quantum,lidar2009adiabatic,rezakhani2010accuracy}. In this paper, we shall use the heuristic condition to inform our choice of interpolation schedule $s(t)$, but we do not rely on it mathematically. By analyzing the algorithm geometrically, we find a rigorous upper bound on the error of the approximation in terms of the schedule $s(t)$. The requirement that this bound on the error be small can be used as a more rigorous and less vague replacement for Eq.~(\ref{eq:adiabaticcondition}).
%
%
\subsection{Geometry of the Algorithm}\label{sec:geometryofalgo}
As in the gate-model analysis of Grover's algorithm, the symmetry of the adiabatic search algorithm allows us to reduce the state space to two dimensions, spanned by $\ket E$ and $\ket {\bar E}$ where $\ket E$ is the equal superposition over the $M$ target states (or ``end" state), and $\ket {\bar E}$ is the equal superposition over the $N-M$ non-target states. We let $\lambda = M/N$ be the fraction of target states allowing us to write $\ket B = \sqrt{1-\lambda} \ket {\bar E} + \sqrt{\lambda} \ket E$. The projector $P$ acts on this subspace like $\ket E \bra E$. As the state evolves by (\ref{eq:schrodinger}) it will always remain in this subspace.  We change basis by letting $\ket 0 = \cos(\mu) \ket {\bar E} + \sin(\mu) \ket E$ and $\ket 1 = -\sin(\mu) \ket {\bar E} + \cos(\mu) \ket E$, with $\cos(2 \mu) = \sqrt{1-\lambda}$. In this basis Eqs.~(\ref{eq:H0}) and (\ref{eq:H1}) can be written as 
\begin{eqnarray}
H_0 &=& \frac{1}{2}I-\frac{1}{2} \hat {n}_0 \cdot \vec{\sigma} \label{eq:H0geo} \\
H_1 &=& \frac{1}{2}I-\frac{1}{2} \hat {n}_1 \cdot \vec{\sigma} \label{eq:H1geo}
\end{eqnarray}
where
\begin{eqnarray}
\hat {n}_0 &=& \sqrt{\lambda} \; \hat x + \sqrt{1-\lambda} \;  \hat z \label{eq:n0}\\
\hat {n}_1 &=& \sqrt{\lambda} \; \hat x - \sqrt{1-\lambda} \; \hat z
\label{eq:n1}
\end{eqnarray}
are normalized vectors and $\vec{\sigma}$ is the vector of Pauli matrices $(X,Y,Z)$. The purpose of the basis change is to make $\hat{n}_0$ and $\hat{n}_1$ symmetric about the $xy$-plane. 

From Eq.~(\ref{eq:H}), we write the Hamiltonian in this 2D space as
\begin{equation}
H(s) = \frac{1}{2}I-\frac{\Delta_\lambda}{2} \hat {n} \cdot \vec{\sigma}
\label{eq:Hgeo}
\end{equation}
where 
\begin{eqnarray}
\Delta_\lambda(s) &=& \sqrt{1-4s(1-s)(1-\lambda)} \label{eq:gap} \\
\hat n(s) &=& \frac{1}{\Delta_\lambda} \Big(\sqrt{\lambda} \; \hat x + \sqrt{1-\lambda}(1-2s) \; \hat z \Big)\label{eq:n}\\
&=& \sin(\theta(s)) \; \hat x + \cos(\theta(s)) \hat z \label{eq:nangle}
\end{eqnarray}
and
\begin{equation}
\theta(s) = \arccos(\hat n \cdot \hat z) = \arccos\Big(\frac{(1-2s)\sqrt{1-\lambda}}{\Delta_\lambda}\Big)
\label{eq:theta}
\end{equation}
is the angle the normalized vector $\hat n$ makes with the $z$-axis.
Using Eqs.~~ (\ref{eq:gap}) and (\ref{eq:theta}), it can be verified that
\begin{equation}
\frac{d\theta}{ds} = \frac{2\sqrt{\lambda(1-\lambda)}}{\Delta_\lambda^2}.
\label{eq:dthetads}
\end{equation}

We will also find it useful to describe states by their position on the Bloch sphere. A normalized vector $\hat r$ can be written as
\begin{equation}
\hat r = \sin(\xi)\cos(\phi) \hat x + \sin(\xi)\sin(\phi)\hat y + \cos(\xi) \hat z
\label{eq:rhatvector}
\end{equation}
The state which is mapped to the point $\hat r$ on the Bloch sphere is
\begin{equation}
\ket {\hat r} = \cos(\xi/2)\ket 0 + e^{i\phi} \sin(\xi/2) \ket 1
\label{eq:rhatstate}
\end{equation}
It is quick to show that
\begin{eqnarray}
\hat r \cdot \vec{\sigma} \ket {\hat r} &=& \ket {\hat r} \\
\hat r \cdot \vec{\sigma} \ket {-\hat r} &=& -\ket {-\hat r}
\label{eq:rhatoperator}
\end{eqnarray}
Thus, referring to Eq.~(\ref{eq:Hgeo}), the eigenvectors of $H(s)$ are $\ket{\hat{n}(s)}$ and $\ket{-\hat{n}(s)}$, and the gap between their eigenvalues is $\Delta_\lambda$. 

\begin{figure}[ht]
\center
\includegraphics[scale=0.33,clip=true,angle=0]{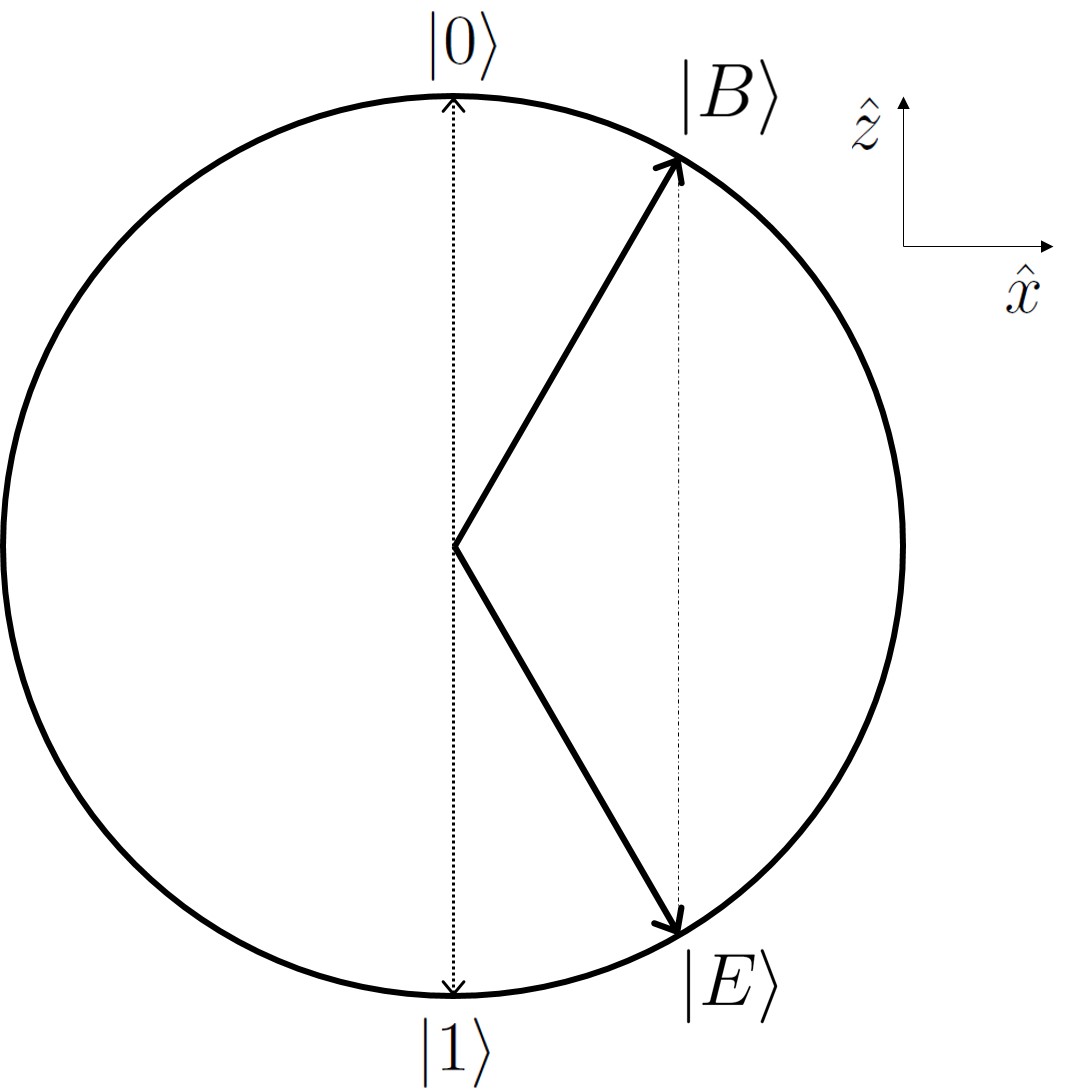}
\caption{Geometric picture of the algorithm. This is the $xz$-plane of the Bloch sphere for $\lambda = 1/4$. The Hamiltonian vector $\Delta_\lambda \hat n$ associated with $H(s)$ follows the dotted line between the begin state and the end state.}
\label{fig:xzplane}
\end{figure}

By viewing the state of the system $\ket{\psi(t)}$ by its position on the Bloch sphere and the Hamiltonian $H(t) = I/2-\vec{v}(t)\cdot\vec{\sigma}/2$ by the ``Hamiltonian vector" $\vec{v}(t)$, it is apparent how we can think the system as a spin-1/2 particle exposed to a magnetic field of strength $\Delta_\lambda$ pointing in the $\hat n$ direction. The tip of the magnetic field vector $\Delta_\lambda \hat n$ associated with the Hamiltonian traces out a vertical line in the $xz$-plane as shown in Figure \ref{fig:xzplane}. With this in mind, we can see intuitively how the algorithm works. The spin of the particle precesses on the Bloch sphere around the magnetic field at a rate proportional to the field strength. At first, the spin and field are aligned in the direction of $\ket B$. The field is rotated from the begin state $\ket B$ to the end state $\ket E$. The precession of the spin around the field causes the field to drag the spin along with it as it changes, so at time $T$ the spin is approximately in the same direction as the field, which corresponds to the desired final state $\ket E$. This is a specific example of the adiabatic theorem in two-dimensions.
%
%
%
%

The gap $\Delta_\lambda(s)$ can be interpreted as the length of the Hamiltonian vector as it changes linearly from $\ket B$ to $\ket E$ in Figure \ref{fig:xzplane}. Notably, $\Delta_\lambda(0)=\Delta_\lambda(1)=1$ and $\Delta_\lambda(1/2)=\sqrt{\lambda}$.

The only piece left to the algorithm is what to choose for the function $s(t)$. Regardless of the form of $s(t)$, the Hamiltonian vector will follow the linear trajectory from the begin state to the end state; the function $s(t)$ determines how fast it moves along this trajectory. We call $s(t)$ the \emph{schedule} of the algorithm. 

Once the schedule has been specified, the final state $\ket {\psi(T)}$ can be computed by evolution via the Schr{\"o}dinger equation (\ref{eq:schrodinger}). For a given schedule $s(t)$, we can compute the probability of measuring a target state at the end of the algorithm as a function of $\lambda$
\begin{equation}\label{eq:successprobability}
P(\lambda) = \lvert \braket{E}{\psi(T)}\rvert^2.
\end{equation}
We also define the \textit{error amplitude} $\delta(\lambda)$ so that $P(\lambda) = 1-\delta(\lambda)^2$. 

We note that when $\lambda = 0$, there are no target states, and when $\lambda = 1$, all the states are target states, so regardless of schedule, $P(0)=0$ and $P(1) = 1$.

In the next section, we discuss how the error amplitude $\delta(\lambda)$ depends on the choice of schedule $s(t)$ and whether the algorithm has the fixed-point property.
%
%
\section{Scheduling and the Fixed-Point Property}\label{sec:schedulingandfixedpoint}

%
%
\subsection{Fixed-point property}\label{secfixedpointsection}
For our analysis, we consider families of schedules $s(t; \epsilon, w)$ parameterized by $\epsilon$ and $w$. The $\epsilon$ parameter is typically small and represents how fast the interpolation occurs. The $w$ parameter represents a lower bound on the fraction of target states $\lambda$. Each schedule family does not depend on $\lambda$ since we assume in general that we do not know $\lambda$ precisely, all we know is the lower bound $w \leq \lambda$. This is a reasonable scenario: if we know that at least one item is marked, then $\lambda \geq 1/N$. Moreover, some problems admit tighter lower bounds, such as the problem of preparing the relatively-prime state, which we discuss in section \ref{sec:applications}. 

Together, $\epsilon$ and $w$ determine the total run time $T$ by requiring $s(0; \epsilon,w)=0$ and $s(T; \epsilon,w)=1$. In section \ref{sec:schedules}, we give three examples of different families of schedules. 

We denote the family of success probability functions which correspond to the family of schedules $s(t;\epsilon,w)$ by $P(\lambda;\epsilon,w)$ and, by extension, the family of error amplitudes by $\delta(\lambda;\epsilon,w)=\sqrt{1-P(\lambda;\epsilon,w)}$. We now formally stipulate conditions on $P(\lambda; \epsilon, w)$ that make the search algorithm under schedule $s(t; \epsilon, w)$ a fixed-point algorithm.

\begin{definition}{Fixed-point property}
The adiabatic search algorithm operating under schedule family $s(t; \epsilon, w)$ has the fixed-point property if there exists a function $f(\epsilon)$ independent of $\lambda$ and $w$ such that $P(\lambda; \epsilon, w) \geq 1-(f(\epsilon))^2$, for all $\lambda \geq w$ and $f(\epsilon) \rightarrow 0$ as $\epsilon \rightarrow 0$. 
\end{definition}

If we have an adiabatic fixed-point search algorithm, and we wish to make $P(\lambda) \geq 1-\delta^2$ for all $\lambda \geq w$, we need only make $\epsilon$ small enough so that $f(\epsilon) \leq \delta$. Thus we view $\epsilon$ and $w$ as independent input parameters which control two features of our success probability function $P(\lambda)$. The parameter $\epsilon$ controls an upper bound on the failure probability $f(\epsilon)^2$, and $w$ controls the range over which this success guarantee is valid. The price paid for a decrease in $w$ or $\epsilon$ is an increase in the run time $T$. 

Although the notion of a schedule and the slowness parameter $\epsilon$ might not make sense in the gate model, the adiabatic fixed-point definition is designed to mirror the definition in the gate model. A gate-model fixed-point algorithm, such as that presented in \cite{yoder2014fixed}, is constructed given parameters $\delta$ and $w$ to succeed with probability $P(\lambda) \geq 1-\delta^2$ so long as $\lambda \geq w$. 

We are also interested in the scaling of the algorithm. The algorithm has Grover-like scaling if the run time $T = \bigoh{1/\sqrt{w}}$.

It is possible for a family of schedules to lack the fixed-point property if $\delta(\lambda; \epsilon, w)$ cannot be bounded by a function only of $\epsilon$. It is also possible for a family of schedules to possess the fixed-point property but lack Grover-like scaling if the dependence of $T$ on $w$ is worse than $\bigoh{1/\sqrt{w}}$. We present examples of these cases in section \ref{sec:schedules}, where we consider three different families of schedules.
%
%
Determining if a certain schedule family has the fixed-point property requires analysis of the error in the adiabatic approximation. The tool we use for this purpose is Theorem \ref{thm:generalbound}, presented in the next section.

\subsection{General bound on error probability}\label{sec:generalbound}
The heuristic condition for validity of the adiabatic theorem in Eq.~(\ref{eq:adiabaticcondition}) is inadequate for several reasons. First, we have not rigorously defended it, and it has actually been shown not to be a sufficient condition for adiabaticity in all cases \cite{marzlin2004inconsistency,tong2005quantitative}. But also, it fails to provide information about the error amplitude $\delta(\lambda)$, besides that it is small. In particular, it does not allow us to understand how our choice of schedule $s(t)$ affects the error amplitude $\delta(\lambda)$. Such information would be practically useful for actually running the adiabatic search algorithm and also allows us to understand the asymptotic behavior of $\delta(\lambda)$ as $T\rightarrow\infty$ and as $\lambda\rightarrow 0$. 

We wish to provide this information by finding an upper bound for $\delta(\lambda)$ in terms of the schedule $s(t)$. As a starting point, we can evaluate Eq.~(\ref{eq:adiabaticcondition}) using the geometric framework we set up in section \ref{sec:geometryofalgo}. From Eq.~(\ref{eq:Hgeo}), we see that the ground and first-excited states of $H(s)$ are $\ket{\hat n}$ and $\ket{-\hat n}$, so the left-hand-side of Eq.~(\ref{eq:adiabaticcondition}) is
\begin{eqnarray}
\lvert\bra{-\hat n}\dot{H}\ket{\hat n}\rvert &=& \dot{s}(t)\lvert\bra{-\hat n}\frac{1}{2}(\hat{n}_0-\hat{n}_1) \cdot \vec{\sigma} \ket{\hat n}\rvert \nonumber \\
&=& \dot{s}(t)\sqrt{1-\lambda}\lvert\bra{-\hat n} Z \ket{\hat n}\rvert \nonumber \\
&=& \dot{s}(t) \frac{\sqrt{\lambda(1-\lambda)}}{\Delta_\lambda}. \label{eq:overlap}
\end{eqnarray}
where our computation relies on Eqs.~(\ref{eq:H}), (\ref{eq:H0geo}-\ref{eq:n1}), and (\ref{eq:rhatstate}). We note that since the minimum of $\Delta_\lambda$ is $\sqrt{\lambda}$, this quantity is at most $\dot{s}(t)$. Since the eigenvalue gap is $\Delta_\lambda$, Eq.~(\ref{eq:adiabaticcondition}) reads
\begin{equation}\label{eq:dsdtbound}
\frac{ds}{dt}  \ll \frac{\Delta_\lambda^3}{\sqrt{\lambda(1-\lambda)}}.
\end{equation}

Now, in Theorem \ref{thm:generalbound}, we present an upper bound on the error amplitude $\delta(\lambda)$ which can be computed given $s(t)$. Thus, the requirement that $\delta(\lambda)$ be small combined with Theorem \ref{thm:generalbound} serves as an alternative adiabatic condition for $s(t)$ which is more concrete (and more rigorous) than Eq.~(\ref{eq:dsdtbound}). 


\begin{maintheorem}\label{thm:generalbound}
If the adiabatic search algorithm is run according to schedule $s(t)$ then  $\delta(\lambda) \leq d_0+d_1$ where 
\begin{eqnarray}
d_0 &=& 2\sqrt{\lambda(1-\lambda)}\dot{s}(0) \label{eq:generalboundd0} \\
d_1 &=& \int_0^T dt \left\lvert \frac{d}{dt}\left(\frac{\sqrt{\lambda(1-\lambda)}}{\Delta_\lambda^3}\frac{ds}{dt}\right)\right\rvert \label{eq:generalboundd1}
\end{eqnarray}
\end{maintheorem}

Theorem \ref{thm:generalbound} is proved in section \ref{sec:pfthmgeneralbound}. The error is related to the derivatives of the schedule; we think of $d_0$ as the initial error due to the choice of derivative at $t=0$ and $d_1$ as the additional error accrued during the algorithm. 
If we require the bound in Theorem \ref{thm:generalbound} on $\delta(\lambda)$ to be small, then $d_0$ and $d_1$ must be small. Since $d_0$ is small, the quantity $\dot{s}\sqrt{\lambda(1-\lambda)}/\Delta_\lambda^3$ must be small for $s=0$ and since $d_1$ is also small, that quantity must remain small for all $s$, implying Eq.~(\ref{eq:dsdtbound}) is satisfied. 

However, the converse relationship is not true. Eq.~(\ref{eq:dsdtbound}) implies that $d_0$ is small but does not necessarily imply that $d_1$ is small. It is possible that $\dot{s}(t)$ is small throughout the algorithm while $\ddot{s}(t)$ is not small. Thus, the condition $d_0+d_1 \ll 1$ is strictly stronger than the heuristic adiabatic condition derived from Eq.~(\ref{eq:adiabaticcondition}).


\subsection{Schedules}\label{sec:schedules}

In this section, we apply Theorem \ref{thm:generalbound} to three families of schedules. Our first two families fail to have both Grover-like scaling and the fixed-point property, although we note how they can be made to have one of the two traits. The third family avoids the issues facing the first two and has both properties.


\subsubsection{Constant-speed schedule}
First, we consider the constant-speed schedule family $s_c(t; \epsilon_c,w)$ defined by $ds_c/dt = \epsilon_c$. This definition does not depend on the $w$ parameter. Boundary conditions imply $T_c = 1/\epsilon_c$, and
\begin{equation}
s_c(t;\epsilon_c,w) =\epsilon_c t = t/T_c.
\label{eq:slin}
\end{equation}
%
%

We can evaluate the bound in Theorem \ref{thm:generalbound} exactly, finding that $d_0 = 2\sqrt{\lambda(1-\lambda)}\epsilon_c$ and $d_1 = 2\sqrt{1-\lambda}(\lambda^{-1}-\sqrt{\lambda})\epsilon_c$. So our bound reads $\delta(\lambda; \epsilon_c,w) \leq 2\epsilon_c\sqrt{1-\lambda}/\lambda$. Letting $\delta$ be the maximum error amplitude for $\lambda \geq w$, our bound says that $\delta = \max_{\lambda \geq w} \delta(\lambda;\epsilon_c,w) < 2\epsilon_c/w$, which diverges as $w$ approaches 0 with $\epsilon_c$ constant. Hence, there is no function $f(\epsilon_c)$ such that our bound guarantees $\delta \leq f(\epsilon_c)$; this suggests the algorithm is not fixed point. Even equipped with a tighter upper bound than the one we present, it cannot be possible to bound $\delta \leq f(\epsilon_c)$ with $\lim_{\epsilon_c \rightarrow 0} f(\epsilon_c)=0$: the error amplitude $\delta(\lambda;\epsilon_c,w)$ has no $w$-dependence, so $\lim_{w\rightarrow 0}\delta \geq \lim_{\lambda\rightarrow 0}\delta(\lambda;\epsilon_c,w)=1$. Thus, the constant-speed schedule does not have the fixed-point property. 

However, we succeed in getting a concrete upper bound on the error probability in terms of both $w$ and $\epsilon_c$, if not $\epsilon_c$ alone. Decreasing $w$ with constant $\epsilon_c$ causes the bound on the error probability to rise, but this effect can be canceled by a roughly proportional decrease in $\epsilon_c$. The fixed-point definition is not satisfied since both parameters must be adjusted to keep the error probability low, but we can use this fact to create a related family of schedules which is fixed point.

Consider the schedule $s_c'(t;\epsilon_c',w)$ defined by $ds_c'/dt = \epsilon_c' w$. The relationship between the primed and unprimed schedules is simply $\epsilon_c = \epsilon_c'w$, so our bound states that $\delta = \max_{\lambda \geq w} \delta(\lambda) < 2\epsilon_c'$. For this schedule, $T'_c = 1/(\epsilon_c' w) = \bigoh{1/(\delta w)}$. It is fixed point but it lacks Grover-like scaling.

%

\subsubsection{Fast schedule}
Our first attempt failed to achieve a quantum speedup. In our second attempt, we make the schedule faster when the gap is large, and slower when the gap is small. The idea of varying the speed according to the gap was first proposed in \cite{roland2002quantum}, but our schedule varies in proportion to the cube of the gap, while that in \cite{roland2002quantum} varies in proportion to the square (and is presented as our third example). Additionally, previous works have not considered these schedules in the context of the fixed-point property and the lower bound $w \leq \lambda$.  

The \textit{fast} schedule family $s_f(t; \epsilon_f,w)$ is defined by
\begin{equation}
\frac{ds_f}{dt} = \epsilon_f\frac{\Delta_w^3}{\sqrt{w(1-w)}}.
\label{eq:fastcondition}
\end{equation}
This schedule is designed to make the left and right sides of Eq.~(\ref{eq:dsdtbound}) proportional by the constant $\epsilon_f$ for all $s_f$ when $\lambda = w$. From this equation and the boundary conditions $s_f(0)=0$, $s_f(T_f)=1$, it can be shown that the total time must be
\begin{equation}
T_f = \frac{\sqrt{1-w}}{\epsilon_f\sqrt{w}}
\label{eq:Tfast}
\end{equation}
and the schedule is 
\begin{equation}
s_f(t; \epsilon_f,w) = \frac{1}{2}-\frac{1}{2}\left(1-\frac{2t}{T_f}\right)\sqrt{\frac{w}{1-(1-\frac{2t}{T_f})^2(1-w)}}.
\label{eq:sfast}
\end{equation}
Since $T_f = \bigoh{1/\sqrt{w}}$, this schedule has Grover-like scaling, a quadratic speedup over the classical search complexity. However, the Grover-like scaling is only useful if the algorithm actually succeeds. We apply the result of Theorem \ref{thm:generalbound} to get an upper bound on the failure probability. Supposing that $\lambda = w$, we calculate $d_0 = 2\epsilon_f$ and $d_1 = 0$, meaning
\begin{equation}\label{eq:fastbound}
\delta(w;\epsilon_f,w) \leq 2\epsilon_f.
\end{equation}
When $\lambda = w$, the error probability is bounded by a function of $\epsilon_f$, and the run time $T_f$ has Grover-like scaling in $w$! To show that the algorithm is fixed point, we would need this to also be true when $\lambda > w$. 

In fact, for the fast schedule, we can modify the arguments for the bound in Theorem \ref{thm:generalbound} to give an \textit{exact} expression for the error amplitude when $\lambda = w$: 


\begin{maintheorem}\label{thm:fastexact}
If the adiabatic search algorithm is run according to the fast schedule given by (\ref{eq:sfast}), then
\begin{equation}\label{eq:deltaf}
\delta(w;\epsilon_f,w) = \frac{2\epsilon_f}{\sqrt{1+4\epsilon_f^2}}\left\lvert\sin\left(\frac{\sqrt{1+4 \epsilon_f^2}}{2\epsilon_f}\phi_w\right) \right\rvert
\end{equation}
where
\begin{equation}
\phi_w = \arctan\Big(\sqrt{\frac{1-w}{w}}\Big).
\label{eq:phiw}
\end{equation}
\end{maintheorem}

Theorem \ref{thm:fastexact} is proved in section \ref{sec:pfthmfastexact}. This result is significant. If the value of $\lambda$ is known, we can run the adiabatic search algorithm with the fast schedule setting parameter $w = \lambda$, and we have an exact analytic expression for the error in the algorithm implying that it has Grover-like scaling in the fraction $\lambda$. 

Varying the speed of the interpolation to be faster when the gap is larger allows us to recover the Grover speedup seen in the circuit model.

But what about when $\lambda$ is unknown? Suppose all we know is that  $\lambda \geq w$. The heuristic condition from Eq.~(\ref{eq:dsdtbound}) now reads 
\begin{equation}\label{eq:fastfixedpoint}
\epsilon_f \ll \frac{\sqrt{w(1-w)}}{\sqrt{\lambda(1-\lambda)}}\frac{\Delta_\lambda^3}{\Delta_w^3}
\end{equation}
This is problematic when $\lambda$ is large, and $s_f$ is close to 0 or 1. For example, fixing $s_f=0$ (so $\Delta_\lambda = \Delta_w = 1$), and $\lambda = 1/2$, we have $\epsilon_f \ll \sqrt{w}$. Looking at Eq.~(\ref{eq:Tfast}), we see that if $\epsilon_f = \bigoh{\sqrt{w}}$ then $T_f = \bigoh{1/w}$, meaning the quadratic speedup is lost.

We confirm this notion with the more concrete bound in Theorem \ref{thm:generalbound}. We evaluate $d_0 = 2\epsilon_f \sqrt{\lambda(1-\lambda)}/\sqrt{w(1-w)}$ and $d_1 = 2\epsilon_f(1-w^{3/2}\lambda^{-3/2})\sqrt{\lambda(1-\lambda)}/\sqrt{w(1-w)}$. In this case, $\delta(\lambda;\epsilon_f,w) \leq d_0+d_1 \leq 4\epsilon_f \sqrt{\lambda(1-\lambda)}/\sqrt{w(1-w)}$ which increases like $\bigoh{\epsilon_f/\sqrt{w}}$ as $w\rightarrow 0$. We are unable to bound the error probability by a function of $\epsilon_f$ alone, suggesting the algorithm is not fixed point. Other techniques could, in principle, yield a tighter bound on the error probability, but numerical simulations of the algorithm confirm that the algorithm lacks the fixed-point property, so any such attempt would suggest the same result.

As in the constant-speed case, the schedule can be modified to become fixed point. Defining $s_f'(t; \epsilon_f', w)$ by the relation $ds_f'/dt = \epsilon_f' \Delta_w^3$ leads to the relationship $T'_f = 1/(\epsilon_f' w)$ and the bound $\delta(\lambda;\epsilon_f',w) \leq 2\epsilon_f'$ whenever $\lambda \geq w$, meaning the primed algorithm is fixed point. In this case, the cost of the fixed-point property is the loss of the Grover-like scaling, since $T'_f = \bigoh{1/w}$. 
%
\subsubsection{Standard schedule}
\begin{figure}
\includegraphics[width=\columnwidth]{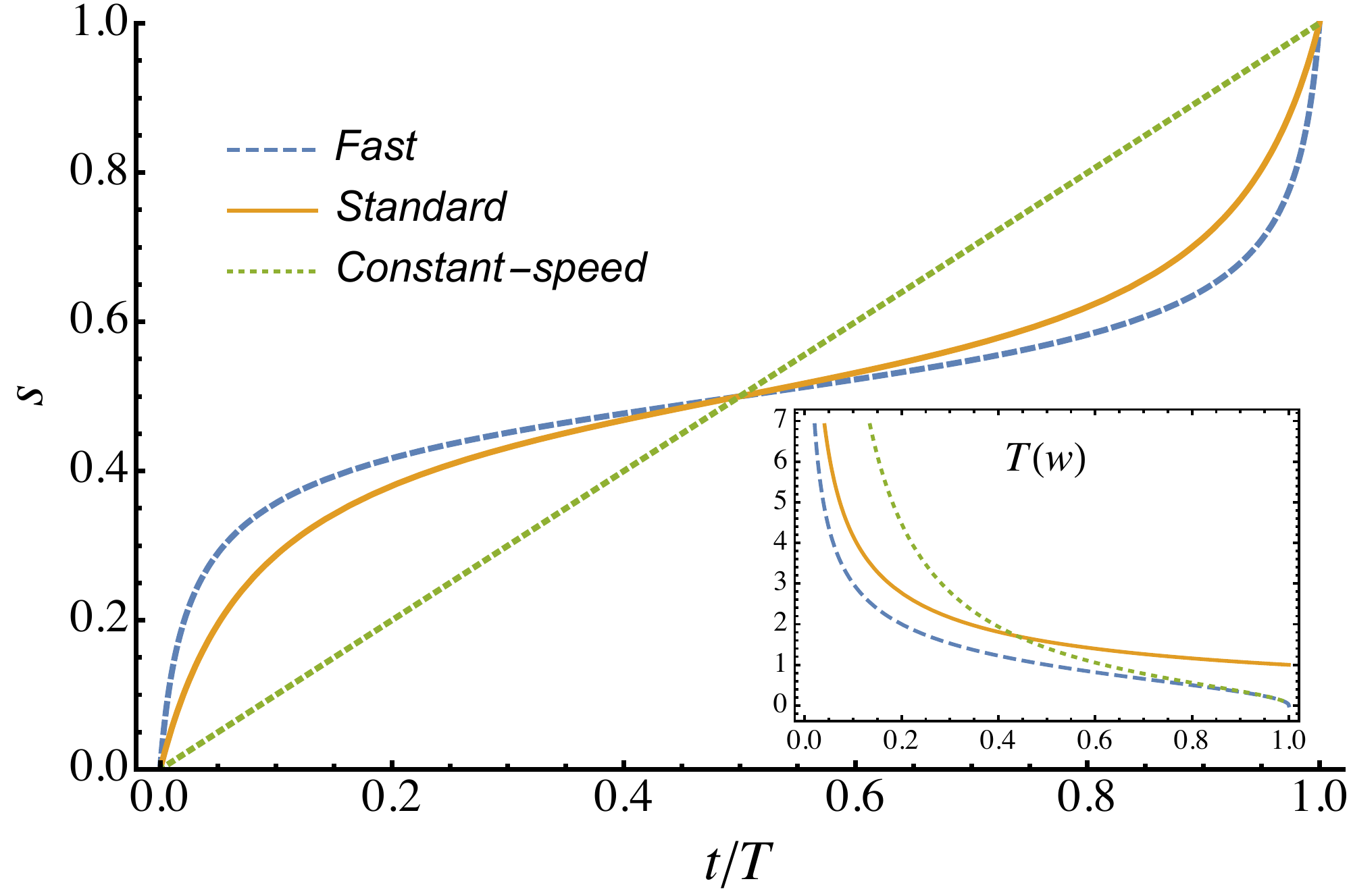}
\caption{\label{fig:sch_compare} A comparison of schedules $s(t)$ at $w=1/20$ plotted with respect to normalized time $t/T$. Both the fast and standard schedules speed up when the gap is large and slow down when it is small, with the fast schedule exhibiting this behavior more strongly. As $w\rightarrow 0$, the fast and standard schedules approach the discontinuous function that is $1/2$ for $t/T\in(0,1)$ and $0$ ($1$) for $t/T=0$ ($1$). As $w\rightarrow1$ both schedules approach the constant-speed schedule. The inset shows the total time $T$ taken (in units of $1/\epsilon$) by these search algorithms as a function of $w$. The fast algorithm is the fastest.}
\end{figure}

The third schedule family we present does not have this problem. The \textit{standard} schedule family $s_s(t;\epsilon_s,w)$, is defined by the relationship
\begin{equation}
\frac{ds_s}{dt}= \epsilon_s \Delta_w^2
\label{eq:dsdt}
\end{equation}
Using the boundary conditions $s_s(0) = 0$, $s_s(T_s) = 1$, it can be shown that 
\begin{equation}
T_s = \frac{\phi_w}{\epsilon_s \sqrt{w(1-w)}}
\label{eq:Tstandard}
\end{equation}
where $\phi_w$ is given by Eq.~(\ref{eq:phiw}), and the schedule is
\begin{equation}
s_s(t; \epsilon_s,w) = \frac{1}{2}-\frac{1}{2} \sqrt{\frac{w}{1-w}}\tan\Big((1-\frac{2t}{T_s})\phi_w\Big)
\label{eq:s}
\end{equation}
By comparison with Eq.~(\ref{eq:Tfast}), we note that the fast schedule is a factor of $\pi/2$ faster than the standard schedule for small $w$ and $\epsilon_s=\epsilon_f$. The three schedules are plotted in Figure \ref{fig:sch_compare}. The standard schedule is so-named because it is the schedule first introduced in \cite{roland2002quantum}. It also appears as the ``constant-norm interpolation" in \cite{rezakhani2010accuracy}.

%
We first examine the heuristic condition from Eq.~(\ref{eq:dsdtbound}), which reads $\epsilon_s \ll \Delta_\lambda^3/(\Delta_w^2 \sqrt{\lambda(1-\lambda)})$, and is satisfied for all $\lambda \geq w$ so long as $\epsilon_s$ is small. This suggests that the error will be small for all $\lambda \geq w$ while the run time complexity is Grover-like $\bigoh{1/\sqrt{w}}$. 

To be more rigorous we apply the bound in Theorem \ref{thm:generalbound}. The computation is not as straightforward as for the other two schedules but is carried out in Appendix A and summarized by Theorem \ref{thm:standardfixedpoint}.


\begin{maintheorem}\label{thm:standardfixedpoint}
If the adiabatic search algorithm is run with parameters $w$ and $\epsilon_s$ according to the standard schedule given by Eq.~(\ref{eq:s}), then
\begin{equation}
\delta = \max_{w\leq \lambda \leq 1} \delta(\lambda;\epsilon_s,w) \leq 2\epsilon_s
\label{eq:continuousdeltabound}
\end{equation}
\end{maintheorem}

Theorem \ref{thm:standardfixedpoint} is unique among previous rigorous treatments of the adiabatic theorem, as it relates to AQC. As mentioned, some previous papers \cite{farhi2000quantum} merely use Eq.~(\ref{eq:adiabaticcondition}) as a heuristic condition for adiabaticity or something similar. Some go further and prove asymptotic statements about the relationship between the error probability and the run time \cite{reichardt2004quantum,avron1987adiabatic}, and some even compute an asymptotic $1/T$ expansion of the error probability \cite{rezakhani2010accuracy,joye1991full,lidar2009adiabatic,hagedorn2002elementary}, but often this series is truncated after a finite number of terms (which is valid only for large $T$). Here we give an explicit upper bound on the error for general schedule in Theorem \ref{thm:generalbound}, and apply this bound to the standard schedule in Theorem \ref{thm:standardfixedpoint}, which confirms asymptotic notions, but is also valid for finite values of the run time $T$ and parameter $w$.

Theorem \ref{thm:standardfixedpoint} rigorously establishes that the standard schedule is a fixed-point algorithm with Grover-like scaling. Since $\delta = \bigoh{\epsilon_s}$,  Eq.~(\ref{eq:Tstandard}) implies that 
\begin{equation}
T_s = \frac{\phi_w}{\bigoh{\delta} \sqrt{w(1-w)}} = \mathcal{O}\Big(\frac{1}{\delta \sqrt{w}}\Big)
\label{eq:Tscaling}
\end{equation}
Comparing this adiabatic algorithm to the fixed-point algorithm in \cite{yoder2014fixed}, which has scaling $\bigoh{\log(2/\delta)/\sqrt{w}}$, we can see that it has similar scaling in $w$, but exponentially worse scaling in $\delta$.

However, if the goal is simply to find the target state, logarithmic scaling in $\delta$ can be achieved by repetition of our algorithm. To see this, suppose we fix $\delta^2=1/3$, and repeat the algorithm $\log(\delta'^2)/\log(1/3)$ times.  The algorithm fails only if each individual run fails, which occurs with probability $\delta'^2$. The total time of this algorithm is logarithmic in the eventual error amplitude $\delta'$: $T'_s = \bigoh{\log(1/\delta')}$. 

Additionally, other schedules are possible which might have better scaling in $\delta$. In particular, it is discussed in \cite{rezakhani2010accuracy} how setting the first $k$ time derivatives of $s(t)$ to zero at the beginning and end of the algorithm makes the first $k$ terms in the $1/T$ expansion of the error amplitude vanish, suggesting that $\delta = \bigoh{1/T^{k+1}}$. If all of the time derivatives vanish at $s=0$ and $s=1$, we might even be able to reproduce the logarithmic dependence on $\delta$ without resorting to repetition. These notions are explored in an asymptotic sense in \cite{rezakhani2010accuracy}, but no bound for finite $T$ and $w$ is given as we do in Theorem \ref{thm:generalbound}.


\section{Geometric intuition and proofs} \label{sec:intuitionandproofs}

Theorems \ref{thm:generalbound} and \ref{thm:fastexact} are proved in this section, while Theorems \ref{thm:standardfixedpoint} and \ref{thm:simulatedgeneralbound} are proved in the appendix. Throughout, when we refer to a Hamiltonian $\mathcal{H}= I/2 - \vec{v}\cdot\vec{\sigma}/2$ in a geometric context, we mean the Hamiltonian vector $\vec v$. 

\begin{definition}{Geometric Intuition} At its core, the method for arriving at Theorems \ref{thm:generalbound} and \ref{thm:fastexact} is very intuitive. We view the Hamiltonian and state vectors geometrically on the Bloch sphere. As the Hamiltonian is changed, the state vector strays away from the Hamiltonian vector. We would like to find how far away it moves, which we quantify by computing an upper bound on the angle between the state and Hamiltonian vectors at the end of the algorithm. We do this by looking at the state from the point of view of the Hamiltonian; we change coordinates so that the Hamiltonian always points in the $\hat z$ direction, and the state is nearly in the $\hat z$ direction. This coordinate transformation is time-dependent, which gives rise to another term in the effective Hamiltonian. For the fast schedule, the direction of the effective Hamiltonian remains fixed, although its magnitude changes. Since the state vector in these coordinates precesses about the effective Hamiltonian, the angle between the state vector and effective Hamiltonian is constant in time, allowing for exact calculation of the error amplitude $\delta(w)$, and leading to Theorem \ref{thm:fastexact}. For general schedule, the direction of the effective Hamiltonian is not constant in time, but it changes very little. As a result, the angle between the state vector and the Hamiltonian changes very little and can be bounded, yielding Theorem \ref{thm:generalbound}. 
\end{definition}

The vital mathematical concept is the time-dependent change of coordinates which rotates the Hamiltonian back to the $\hat z$ direction, and is described by Lemma \ref{lem:effectiveH}. The change of coordinates is defined by the state
\begin{equation}
\ket {\phi(t)} = \exp( i \, \frac{\theta}{2}\, Y ) \ket {\psi(t)}
\label{eq:ketphi}
\end{equation}
where $\theta$ is given by Eq.~(\ref{eq:theta}). At any instant of time, the relationship between $\ket \psi$ and $\ket \phi$ is simply a rotation of the Bloch sphere by angle $-\theta$ about the $\hat y$-axis.


\begin{lemma}\label{lem:effectiveH}
$\ket {\phi(t)}$ evolves according to the Schr{\"o}dinger equation 
\begin{equation}
i\frac{d}{dt} \ket {\phi(t)} = H_\phi \ket{\phi(t)}
\label{eq:schrodingerphi}
\end{equation}
where $H_\phi$ is the effective Hamiltonian given by
\begin{equation}
H_\phi = \frac{1}{2}I - \frac{m}{2}\hat{n}_\phi \cdot \vec{\sigma}
\label{eq:Hphi}
\end{equation}
with
\begin{equation}
m = \sqrt{\Delta^2_\lambda + (\dot \theta)^2}
\label{eq:m}
\end{equation}
and 
\begin{equation}
\hat n_\phi = \frac{1}{m}(\dot \theta \hat y +  \Delta_\lambda \hat z)
\label{eq:nphi}
\end{equation}
\end{lemma}


\begin{mainproof}{Proof of Lemma \ref{lem:effectiveH}}
This lemma follows from application of the Schr{\"o}dinger equation (\ref{eq:schrodinger}) on the definition of $\ket \phi$ in Eq.~(\ref{eq:ketphi}).
\begin{equation} 
i\frac{d}{dt} \ket {\phi(t)} = \Big(-\frac{\dot \theta}{2} Y + \exp(i\, \frac{\theta}{2}\, Y)H \exp(-i\, \frac{\theta}{2}\, Y) \Big)\ket{\phi(t)} \nonumber
\end{equation}
Using the geometric expression (\ref{eq:Hgeo}) for $H$, we have
\begin{eqnarray} 
&&\exp(i\, \frac{\theta}{2}\, Y)H \exp(-i\, \frac{\theta}{2}\, Y) \nonumber \\
&=& \frac{1}{2}I- \frac{\Delta_\lambda}{2}\exp(i\, \frac{\theta}{2}\, Y)(\hat n \cdot \sigma) \exp(-i\, \frac{\theta}{2}\, Y) \nonumber \\
&=& \frac{1}{2}I-\frac{\Delta_\lambda}{2}Z \nonumber
\end{eqnarray}
and thus, 
\begin{equation}
H_\phi = \frac{1}{2}I-\frac{\Delta_\lambda}{2}Z-\frac{\dot \theta}{2} Y
\nonumber
\end{equation}
and the lemma is proved.
\end{mainproof}


The term $\Delta_\lambda Z/2$ in $H_\phi$ comes from rotation of the Hamiltonian back to the $\hat z$-axis. The term $\dot{\theta}Y/2$ comes from the time-dependence of this rotation.


\subsection{Proof of Theorem \ref{thm:generalbound}} \label{sec:pfthmgeneralbound}

\begin{figure}[htbp]
\subfloat[]{\includegraphics[height=0.7in]{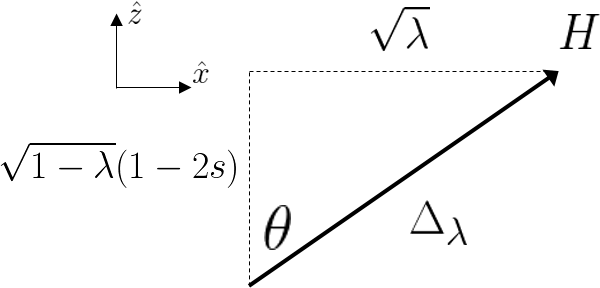}}
\hspace{12 pt}
\subfloat[]{\includegraphics[height=0.7in]{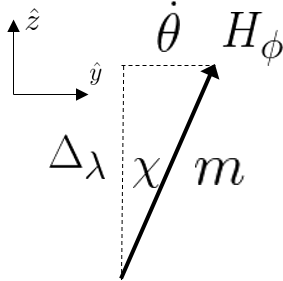}}
\hspace{12 pt}
\subfloat[]{\includegraphics[height=0.7in]{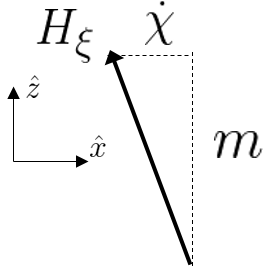}}
\caption{\label{fig:prooffig} Geometric guide to the proof of Theorem \ref{thm:generalbound}. We perform two time-dependent coordinate changes. (a) The state $\ket \psi$ evolves by Hamiltonian $H$, consistent with Figure \ref{fig:xzplane}. (b) The state $\ket \phi$ is related to $\ket \psi$ by Eq.~(\ref{eq:ketphi}) and evolves by Hamiltonian $H_\phi$. (c) The state $\ket \xi$ is related to $\ket \phi$ by Eq.~(\ref{eq:ketxi}) and evolves by Hamiltonian $H_\xi$. Each Hamiltonian $\mathcal{H}= I/2 - \vec{v}\cdot\vec{\sigma}/2$ is represented by the vector $\vec{v}$ in the figure.}
\end{figure}

To prove Theorem \ref{thm:generalbound}, we will make two coordinate changes and use Lemma \ref{lem:effectiveH} twice. The effective Hamiltonian vectors in each frame are depicted graphically in Figure \ref{fig:prooffig}. Building off of Lemma \ref{lem:effectiveH}, we define

\begin{equation}\label{eq:chi}
\chi(t) = \arctan(\dot{\theta}(t)/\Delta_\lambda(t))
\end{equation}
to be the angle between $\hat n_\phi$ and $\hat z$, allowing us to rewrite (\ref{eq:nphi}) as
\begin{equation}\label{eq:nphiangle}
\hat{n}_{\phi} = \sin(\chi) \hat{y} + \cos(\chi) \hat{z},
\end{equation}
noting the similarity to Eq.~(\ref{eq:nangle}). If the interpolation is slow, $\dot{\theta}$, and hence $\chi$ will be small, so $\hat{n}_{\phi}$ will be close to $\hat{z}$. 

Now, we repeat the process again, this time using $\hat n_\phi$ in place of $\hat n$ and rotating in the $yz$-plane instead of the $xz$-plane. We let
\begin{equation}
\ket {\xi(t)} = \exp(- i \, \frac{\chi}{2}\, X ) \ket {\phi(t)}
\label{eq:ketxi}
\end{equation}
and by invoking the result from Lemma \ref{lem:effectiveH} (except in the $yz$-plane instead of $xy$-plane), we see that $\ket {\xi(t)}$ evolves according to 
\begin{equation}
i\frac{d}{dt} \ket {\xi(t)} = H_\xi \ket {\xi(t)}
\label{eq:schrodingerxi}
\end{equation}
with 
\begin{equation}
H_\xi = \frac{1}{2}I - \frac{1}{2}\hat n_\xi \cdot \vec{\sigma} = \frac{1}{2}I-\frac{m}{2} Z + \frac{\dot \chi}{2} X
\label{eq:Hxi}
\end{equation}
We note that $\ket {\xi(0)} = \cos(\chi(0)/2) \ket 0 - i\sin(\chi(0)/2) \ket 1$. The most general possible equation for $\ket {\xi(t)}$ is 
\begin{equation}
\ket {\xi(t)} = e^{iq(t)}\left(\cos\Big(\frac{r(t)}{2}\Big) \ket 0 + e^{ip(t)} \sin\Big(\frac{r(t)}{2}\Big) \ket 1 \right).
\label{eq:ketxigeneral}
\end{equation}
Note that $r$ is the angle between $\hat z$ and $\ket \xi$ on the Bloch sphere, but since $\ket \xi$ and $\ket \phi$ are related by the rotation (\ref{eq:ketxi}), which preserves angles, $r$ is also the angle between the direction of the effective Hamiltonian $H_\phi$ and the state $\ket \phi$. 

By plugging (\ref{eq:ketxigeneral}) into the Schr{\"o}dinger equation (\ref{eq:schrodingerxi}), we can prove the following:


\begin{lemma}\label{lem:drdt}
$\lvert \dot{r} \rvert \leq \lvert \dot{\chi} \rvert$. 
\end{lemma}



\begin{mainproof}{Proof of Lemma \ref{lem:drdt}} Applying the left-hand-side of Eq.~(\ref{eq:schrodingerxi}) to Eq.~(\ref{eq:ketxigeneral}) gives
\begin{eqnarray}
&&e^{iq}\left(-\dot{q}\cos\Big(\frac{r}{2}\Big)-\frac{i}{2}\dot{r}\sin\Big(\frac{r}{2}\Big)\right)\ket 0 +  \nonumber\\
&&e^{i(q+p)}\left(-\dot{q}\sin \Big(\frac{r}{2}\Big)+\frac{i}{2}\dot{r}\cos\Big(\frac{r}{2}\Big) -\dot{p} \sin \Big(\frac{r}{2}\Big)\right)\ket 1
\label{eq:lhs} \nonumber
\end{eqnarray}
and applying the right-hand-side gives
\begin{eqnarray}
&&\frac{1}{2}e^{iq}\left(\cos\Big(\frac{r}{2}\Big)-m\cos\Big(\frac{r}{2}\Big)+e^{ip}\dot{\chi}\sin\Big(\frac{r}{2}\Big)\right)\ket 0 + \nonumber\\
&& \frac{1}{2}e^{iq}\left(e^{ip}\sin\Big(\frac{r}{2}\Big)+\dot{\chi}\cos \Big(\frac{r}{2}\Big)+e^{ip}m\sin\Big(\frac{r}{2}\Big)\right)\ket 1
\label{eq:rhs}. \nonumber
\end{eqnarray}
These quantities must be equal. We can immediately cancel the $e^{iq}$ factor. Now, the imaginary parts of the coefficient of $\ket 0$ for both the left and right-hand-sides must be equal, giving the equation:
\begin{eqnarray}
-\frac{1}{2}\dot{r}\sin\Big(\frac{r}{2}\Big) &=& \frac{1}{2} \dot{\chi} \sin(p)  \sin\Big(\frac{r}{2}\Big) \nonumber \\
\dot{r} &=& -\dot{\chi} \sin(p) \nonumber \\
\lvert \dot{r} \rvert &\leq & \lvert \dot{\chi} \rvert
\label{eq:lhsrhs}
\end{eqnarray}
which proves the lemma.
\end{mainproof}


Physically, Lemma \ref{lem:drdt} is telling us that the angle $r$ between $\ket \phi$ and the effective Hamiltonian $H_\phi$ cannot change faster than the angle $\chi$ between the effective Hamiltonian $H_\phi$ and $\hat z$. This fact could have been proved using a purely geometric argument. 

The success probability (\ref{eq:successprobability}) can be expressed in our new coordinates using the relationships (\ref{eq:ketphi}) and (\ref{eq:ketxi}). 
\begin{equation}
P(\lambda) = \lvert \braket{0}{\phi(T)} \rvert^2 =  \lvert \braket{\xi(0)}{\xi(T)} \rvert^2
\label{eq:Pxi}
\end{equation}
We define
\begin{equation}
A = \frac{r(0)+r(T)}{2}
\end{equation}
Using Eq.~(\ref{eq:ketxigeneral}), we can see that
\begin{eqnarray}
P(\lambda) = \lvert \braket{\xi(0)}{\xi(T)} \rvert^2 \geq \cos^2(A) \geq 1-A^2
\label{eq:PboundA}
\end{eqnarray}
Since $\int_0^T \dot r dt = r(T)-r(0)$, we have 
\begin{eqnarray}
A &=& r(0)+\frac{1}{2}\int_0^T \dot r dt \nonumber \\
&\leq & \chi(0) + \frac{1}{2}\int_0^T \lvert \dot \chi \rvert dt.
\label{eq:Aint}
\end{eqnarray}
Plugging Eq.~(\ref{eq:dthetads}) into the definition of $\chi$ in Eq.~(\ref{eq:chi}), we see that $\chi(t) = \arctan(2\sqrt{\lambda(1-\lambda)}\dot{s}(t)/\Delta_\lambda(t)^3)$. However, when $u>0$, $\arctan(u) \leq u$ and $\lvert d(\arctan(u))/dt \rvert = \lvert \dot u / (1+u^2)\rvert \leq \lvert \dot u \rvert$. Thus, $\chi(0) \leq d_0$ and $\int \lvert \dot{\chi} \rvert / 2 \leq d_1$ where $d_0$ and $d_1$ are given by Eqs.~(\ref{eq:generalboundd0}) and  (\ref{eq:generalboundd1}). Since $\delta(\lambda) = \sqrt{1-P(\lambda)} \leq A$, this proves the theorem.


\subsection{Proof of Theorem \ref{thm:fastexact}}\label{sec:pfthmfastexact}

Theorem \ref{thm:fastexact} gives an exact expression for the error probability where Theorem \ref{thm:generalbound} only gives an upper bound. The structure of the proof is the same as that for Theorem \ref{thm:generalbound}.

We define $\chi(t)$ as in Eq.~(\ref{eq:chi}) for the fast schedule with $\lambda = w$, but as can be verified by Eqs.~(\ref{eq:dthetads}) and (\ref{eq:fastcondition}), $\chi = 2\epsilon_f$ is constant throughout time. So $\hat{n}_\phi = \sin(\chi)\hat{y} + \cos(\chi)\hat{z}$ is fixed. In this frame, the effective Hamiltonian does not change direction, though its strength varies and is given by $m(t) = (\Delta_w^2 + (\dot{\theta})^2)^{1/2} = \Delta_w(1+4\epsilon_f^2)^{1/2}$. Thus, the angle between the state vector and $\hat{n}_\phi$ should remain constant throughout the evolution. Mathematically, this follows from Lemma \ref{lem:drdt}, since $\dot \chi = 0$ implies angle $r$ is constant. It is this fact that allows for exact treatment of the error probability. This is not a coincidence. The proportionality between the gap $\Delta_w$ and the angular velocity $\dot{\theta}$ of $H(t)$ is intimately related to the adiabatic condition in Eq.~(\ref{eq:adiabaticcondition}) and the definition of the fast schedule in Eq.~(\ref{eq:fastcondition}). We define $\ket{\xi}$ as in Eq.~(\ref{eq:ketxi}) but since $\chi$ is constant the coordinate change is no longer time-dependent. We see that $H_\xi = I/2-m(t)Z/2$. 

At $t=0$ we have $\ket{\xi(0)} = \cos(\chi/2)\ket 0 - i \sin(\chi/2)\ket 1$. Evolution by $H_\xi(t)$ simply applies a relative phase on the $\ket 1$ component of $\ket \xi$, so up to a global phase, $\ket {\xi(T_f)} = \cos(\chi/2)\ket 0 - i\exp(-i\int_0^{T_f} m(t)dt) \sin(\chi/2)\ket 1$.

The phase given by the integral can be evaluated as 
\begin{eqnarray}\label{eq:phaseintegral}
\int_0^{T_f} m(t)dt &=& \frac{1}{\epsilon_f}\sqrt{1+4\epsilon_f^2}\sqrt{w(1-w)}\int_0^1 \frac{ds_f}{\Delta_w^2} \nonumber \\
&=& \sqrt{1+4\epsilon_f^2}\frac{\phi_w}{\epsilon_f} 
\end{eqnarray}
where, incidentally, the integral is the same one used to arrive at Eq.~(\ref{eq:Tstandard}), with $\phi_w$ given by Eq.~(\ref{eq:phiw}). Finally, we can evaluate the success probability $P(w; \epsilon_f,w) = \lvert \braket{\xi(0)}{\xi({T_f})}\rvert^2$. We let the integral in Eq.~(\ref{eq:phaseintegral}) be denoted by $p$ and evaluate
\begin{eqnarray}
P(w;\epsilon_f,w) &=& \lvert \cos^2(\chi/2) + \exp(-ip) \sin^2(\chi/2)\rvert^2 \nonumber \\
&=& 1-\sin^2(\chi)\sin^2(p/2) \label{eq:fastsuccess}
\end{eqnarray}
thus $\delta(w;\epsilon_f,w) = \sin(\chi)\lvert\sin(p/2)\rvert$ where $\sin(\chi)=2\epsilon_f / (1+4\epsilon_f^2)^{1/2}$, so the theorem has been proved. 


\section{Simulation in the Gate Model}\label{sec:simulation}

What is the relationship between the standard-schedule adiabatic fixed-point algorithm and the gate-model fixed-point algorithms from \cite{yoder2014fixed}? They bear little resemblance beyond the fixed-point property. The evolution of the state in the adiabatic case is smooth and stays nearly in the $xz$-plane of the Bloch sphere throughout the evolution, while the gate-model algorithm subjects the state to discrete rotations through large angles causing it to jump around all parts of the Bloch sphere. It seems unlikely that this latter sort of evolution could be exhibited by any adiabatic algorithm, but perhaps the former could be exhibited by a gate-model algorithm. Fortunately, simulating the adiabatic search algorithm in the gate-model is simply done using a Trotter formula, as we explain in this section. We adapt our geometric analysis to show that the resulting gate-model algorithm can be made to have the fixed-point property, and thus act as an alternative to the algorithm in \cite{yoder2014fixed} with evolution more closely resembling that of the adiabatic search algorithm itself.

We will do this simulation in two steps: first, we will discretize the continuous Hamiltonian $H(t)$, so it is constant over small time intervals $\delta t$, then we will simulate the discrete Hamiltonian using a Trotter formula. Suppose we divide the total time $T$ into $l$ intervals of width $\delta t = T/l$. Then we let $t_j = j \: \delta t$, and $s_j = s(t_j)$ for some schedule $s(t)$. The discretized Hamiltonian is
\begin{equation}
H_d(t) = H(\delta t \lfloor t/ \delta t \rfloor)
\label{eq:Hd}.
\end{equation}
Thus, $H_d$ agrees with $H$ whenever $t=t_j$ for some $j$, and $H_d$ is constant between $t_j$ and $t_{j+1}$ for all $j$.

\begin{figure}[ht]
\center
\includegraphics[width=\columnwidth]{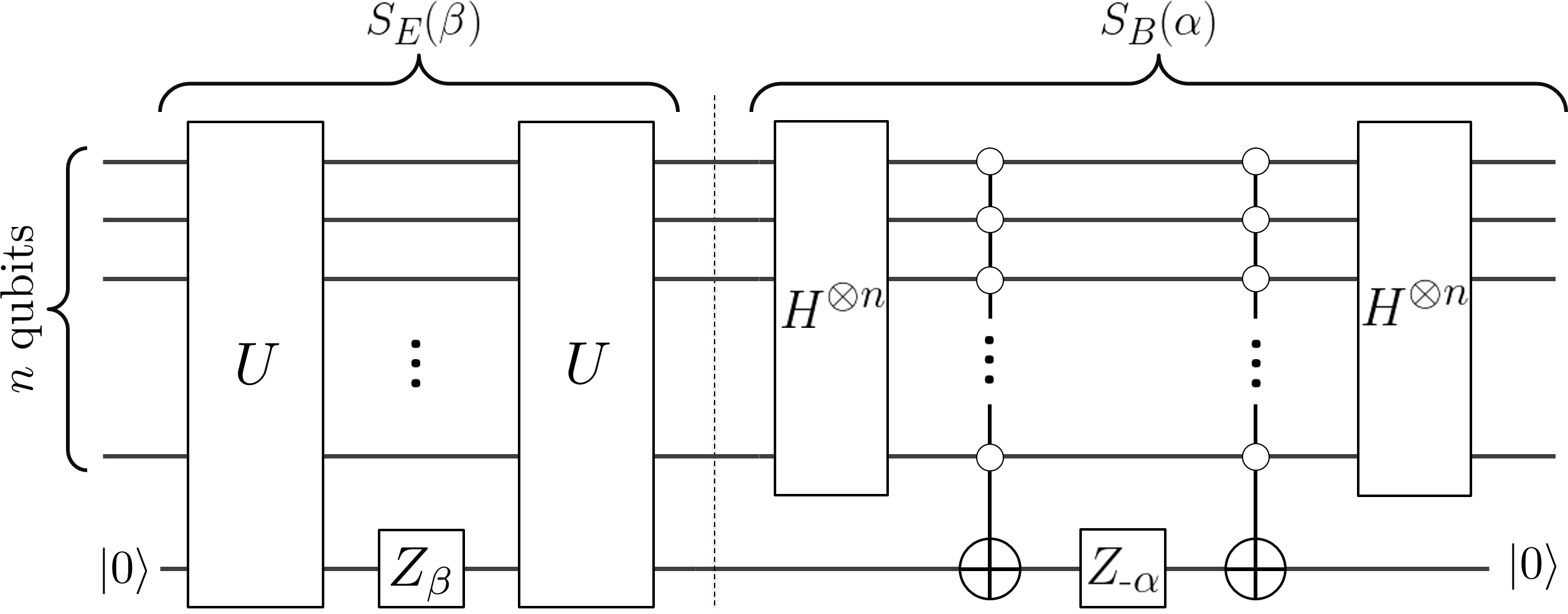}
\caption{The circuit which simulates the discretized Hamiltonian $H_d$ over one time interval of width $\delta t$, performing the unitary operation $U_t^{(j)}$. The gate $Z_\zeta$ applies the operation $\exp(-i\zeta Z/2)$.}
\label{fig:circuit}
\end{figure}

For a given interval $j$, the effect of applying the Hamiltonian $H_d$ is the unitary operator
\begin{eqnarray}
U_d^{(j)} &=& \exp(-i H(t_j)\, \delta t) \\
&=& \exp\Big(-i\, (1-s_j)\, H_0 \,\delta t - i\: s_j \,H_1 \,\delta t \Big).
\label{eq:Ud}
\end{eqnarray}
If $\delta t$ is small, then we can use a Trotter formula to approximate this unitary as a product of two unitaries:
\begin{equation}
U_t^{(j)} = \exp(-i\, (1-s_j)\, H_0 \,\delta t) \exp( - i\: s_j \,H_1 \,\delta t ).
\label{eq:Utexp}
\end{equation}
It can be shown that $U^{(j)}_d(t) = U^{(j)}_t(t) + \mathcal{O}(\delta t^2)$ \cite{nielsen2010quantum}. The benefit of $U_t$ is that it is a product of two unitaries which are easy to implement in a circuit. Exponentiating $H_0$ is equivalent to a rotation about the begin state in the Bloch sphere framework we introduced above. Likewise, exponentiating $H_1$ is a rotation about the end state. Specifically, we have:
\begin{equation}
U_t^{(j)} = S_B(\alpha_j)S_E(\beta_j)
\label{eq:Utproduct}
\end{equation}
up to an overall phase, where
\begin{eqnarray}
S_B(\alpha) &=& I-(1-e^{-i\alpha})\ket B \bra B \nonumber \\
&=& \exp(-i\alpha)\exp(i\alpha H_0)
\label{eq:SB} \\
S_E(\beta) &=& I-(1-e^{+i\beta})\ket E \bra E \nonumber \\
&=& \exp(i\beta)\exp(-i\beta H_1)
\label{eq:SE}
\end{eqnarray}
and $\alpha_j =-(1-s_j)\delta t$ and $\beta_j = s_j \delta t$. The circuit that implements $U^{(j)}_t$ is the same one discussed in \cite{yoder2014fixed} and is shown in Figure \ref{fig:circuit}. It makes use of the discrete oracle $U$, which flips an ancilla bit when fed a target state: $U\ket x \ket 0 = \ket x \ket 1$ if $\ket x$ is a target state, and otherwise acts as the identity. As an example, suppose we are using the standard schedule $s_s(t;\epsilon_s,w)$ given by Eq.~(\ref{eq:s}), then, recalling that we have split $T_s$ into $l$ equal time steps of size $\delta t$, we can say that 
\begin{eqnarray}
\alpha_j =& -&\frac{\delta t}{2}-\frac{\delta t}{2} \sqrt{\frac{w}{1-w}}\tan\Big((1-\frac{2j}{l})\phi_w\Big) \label{eq:alpha} \\
\beta_j =& &\frac{\delta t}{2}-\frac{\delta t}{2} \sqrt{\frac{w}{1-w}}\tan\Big((1-\frac{2j}{l})\phi_w\Big)
\label{eq:beta}
\end{eqnarray}
%
%

Now we can formally define the gate-model algorithm:


\begin{definition}{Simulated Adiabatic Search Algorithm}

\noindent \textbf{Inputs:} schedule $s(t)$ (where $s(0)=0$, $s(T)=1$), $\delta t$ \\
\noindent \textbf{Procedure:} Prepare the initial state $\ket {\psi_0}$ to be $\ket B$. Apply circuit $U_t^{(j)}$ to $\ket {\psi_j}$ to yield $\ket {\psi_{j+1}}$ for $j$ = 0, 1, 2, $\ldots$, $l-1$, where $l = \lfloor T/\delta t \rfloor$.  Measure the system in the computational basis. \\
\noindent \textbf{Output:} The post-measurement state is a target state with probability $P(\lambda) \equiv \lvert\braket{\psi_l}{E}\rvert^2$, where $\lambda$ is the fraction of target states. \\
\end{definition}

We claim that the simulated adiabatic search algorithm, using the standard schedule $s_s(t; \epsilon_s, w)$ given by Eq.~(\ref{eq:s}), is also fixed point with Grover-like scaling. We prove this by bounding the error in the simulated adiabatic search algorithm for general schedule $s(t)$ and then evaluating this bound for the standard schedule (recall $\delta(\lambda) = \sqrt{1-P(\lambda)}$). 
%


\begin{maintheorem}\label{thm:simulatedgeneralbound}
If the simulated adiabatic search algorithm is run according to schedule $s(t)$ and step size $\delta t$, then
\begin{equation}
\delta(\lambda) < 3.1\sqrt{\delta t}+(d_0+d_1)(1+ \delta t^2/25)
\label{eq:discretedeltabound}
\end{equation}
where $d_0$ and $d_1$ are functions of $\lambda$ given by Eqs.~(\ref{eq:generalboundd0}) and (\ref{eq:generalboundd1}) from Theorem \ref{thm:generalbound}.
\end{maintheorem}

Theorem \ref{thm:simulatedgeneralbound} is proved in Appendix B. As $\delta t$ approaches 0, this bound approaches the bound in Theorem \ref{thm:generalbound}. If $s(t)$ is the standard schedule (\ref{eq:s}), then Theorem \ref{thm:standardfixedpoint} says that $d_0+d_1 \leq 2 \epsilon_s$, so for the simulated adiabatic search algorithm with standard schedule, $P(\lambda) \geq 1-\delta^2$ with $\delta = 3.1\sqrt{\delta t}+2\epsilon_s(1+\delta t^2/25)$. The number of oracle queries associated with this sequence of length $l$ is $L = 2l+1$ and $l = \lfloor T/\delta t \rfloor$. Thus, we can express the query complexity in terms of the desired values for $w$ and $\delta$: 
\begin{equation}
L = 1+2\lfloor T_s/\delta t\rfloor = 1+ \frac{2\phi_w}{\epsilon \, \delta t \sqrt{w(1-w)}} = \mathcal{O}\Big(\frac{1}{\delta^3 \sqrt{w}}\Big).
\label{eq:Lscaling}
\end{equation}

Thus, Theorem \ref{thm:simulatedgeneralbound} shows that the simulated adiabatic search algorithm with standard schedule is a fixed-point algorithm which shows Grover's quadratic speedup. This gives an alternative sequence of $\alpha$ and $\beta$ angles (see Eqs.~(\ref{eq:alpha}) and (\ref{eq:beta})) that can be used in place of the fixed-point sequence given in \cite{yoder2014fixed}. It is known that adiabatic computation can be simulated with polynomial-overhead \cite{van2001powerful}, but this algorithm shows that the adiabatic search algorithm can be simulated while maintaining the same scaling in $w$.

We briefly mention an overview of how Theorem \ref{thm:simulatedgeneralbound} is proved. One approach would be to bound the error introduced by discretization (the difference between applying the true Hamiltonian $H$ and the discrete Hamiltonian $H_d$), and then bound the error induced by Trotterization (the difference between applying $U_d^{(j)}$ and $U_t^{(j)}$). We did not find this approach successful for the following reason: suppose the algorithm is run for time $T$ with $l$ time steps, so $\delta t = T/l$. The Trotterization error is $\bigoh{\delta t^2}$ per time step, making the total error $\bigoh{l \delta t^2} = \bigoh{T\delta t}$. Since $T = \bigoh{1/\sqrt{\lambda}}$ and the error must be $\bigoh{1}$, $\delta t = \bigoh{\sqrt{\lambda}}$, and thus the number of time steps $l = \bigoh{1/\lambda}$, which does not have Grover-like scaling. The discretization error is not problematic in this way.

Instead, our proof finds an alternate continuous Hamiltonian $H_t$ which, upon discretization, generates the exact evolution operations $U_t^{(j)}$ applied in the simulated adiabatic search algorithm and given by Eq.~(\ref{eq:Utexp}). This Hamiltonian $H_t$, which is approximately equal to $H$, follows a slightly modified but still continuous path from the begin state $\ket B$ to the end state $\ket E$. In particular, this path leaves the $xz$-plane. We use the same strategy as in the proof from Theorem \ref{thm:generalbound} to bound the error induced by the continuous algorithm defined by application of $H_t$. This error \textit{does} have Grover-like scaling. Then, we discretize $H_t$, giving $H_{td}$, and bound the error induced by discretization. 

While the first approach views Trotterization as an error-inducing step and tries to find an upper bound on this error, the second approach capitalizes on the fact that the Trotterization error at each time step is correlated in a way that can be interpreted as simply an alternate path from $\ket B$ to $\ket E$ on the Bloch sphere and hence should also be a valid adiabatic algorithm. As $T$ gets longer with constant $\delta t$, the first approach finds that the error accrues while the second approach does not. The alternate path is a function only of $\delta t$ and not of the schedule $s(t)$ (which defines how fast the path is traversed) and hence a longer run time does not result in more error due to Trotterization under this approach.


\section{Applications of the fixed-point property}\label{sec:applications}

In this section, we elaborate on two direct applications of fixed-point algorithms: preparation of the relatively-prime state and oblivious amplitude amplification. Other applications are mentioned briefly in the introduction.

\subsection{Preparation of the relatively-prime state}
For a specific example of a problem which might benefit from a fixed-point algorithm, we look to number theory. Given an integer $J$, we wish to prepare the relatively-prime state, the superposition over states which are relatively prime to $J$. 
\begin{equation}\label{eq:relprime}
\ket \Phi = \frac{1}{\sqrt{\phi(J)}}\sum_{\gcd(x,J)=1} \ket x
\end{equation}
where $\phi$ is the Euler totient function. That is, $\phi(J)$ is the number of integers relatively prime to $J$. This can be done by beginning in the equal superposition $\ket B$, running the classical Euclidean algorithm (which is efficient) for calculation of the greatest common divisor (gcd) while in superposition, and copying the result to an ancilla register
\begin{equation}\label{eq:euclideanalgo}
\ket B \ket 0 \rightarrow \sum_{x=0}^{J-1}\ket x \ket {\gcd(x,J)}
\end{equation}
Then, measurement of the ancilla register yields outcome 1 and the state $\ket \Phi$ with probability $\phi(J)/J$. The process is repeated until a 1 outcome is obtained. The algorithm is efficient because there is a lower bound \cite{barkley1962approximate} on $\phi(J)/J$
\begin{equation}\label{eq:philowerbound}
\frac{\phi(J)}{J} >\frac{1}{e^\gamma \log \log J + \frac{3}{\log \log J}}
\end{equation}
where $\gamma \approx 0.577$ is the Euler-Mascheroni constant. so the state $\ket \Phi$ will be produced with high probability after only $\bigoh{\log\log J}$ repetitions of the Euclidean algorithm. 

We can reformulate the algorithm as a search algorithm, with $\lambda = \phi(J)/J$ and the Euclidean algorithm playing the role of the oracle. In the gate model, the oracle consists of doing the euclidean algorithm and flipping an ancilla bit if the gcd is 1. In the adiabatic model, the oracle is a Hamiltonian $H_J$ such that $H_J\ket x = 0$ if $\gcd(x,J) = 1$ and $H_J\ket x = \ket x$ otherwise. Thus, the ground state of $H_J$ is $\ket \Phi$. 

Running a fixed-point search algorithm with lower bound parameter $w$ given by Eq.~(\ref{eq:philowerbound}) will prepare the superposition of target states $\ket \Phi$ with high fidelity using only $\bigoh{\sqrt{\log \log J}}$ queries to the Euclidean algorithm (or total run time in the adiabatic model). This is a quadratic speedup over the algorithm in \cite{love2014efficient} which does not make use of a quantum search algorithm. On an adiabatic computer, this would entail running the adiabatic search algorithm using the standard schedule. Moreover, the fixed-point algorithm allows the state $\ket \Phi$ to be prepared to arbitrary fidelity without the need for measurement. Crucially, we only need to know the lower bound in Eq.~(\ref{eq:philowerbound}) for $\lambda=\phi(J)/J$ for the algorithm to work.

\subsection{Fixed-point oblivious amplitude amplification}
Oblivious amplitude amplification \cite{berry2014exponential} is a technique for implementing, on a state $\ket{\psi}$, a desired unitary $V$ that we cannot construct directly with a larger unitary $U$ that we can construct, at least when given some ancilla qubits. The ``oblivious'' in the name comes from the fact that amplitude amplification still works despite not having the full ability to reflect about the initial state $\ket{\psi}$. Thus, this procedure is particularly useful when the initial state $\ket{\psi}$ is not only unknown but also not renewable --- we have only one copy and cannot or would prefer not to make another. For instance, this is the case in Hamiltonian simulation \cite{berry2014exponential}.

For convenience, we consider $|\psi\rangle$ to be an $n$-qubit state and the extension to Hilbert space to be $m$-qubits. Then, write the start state as \cite{berry2014exponential}
\begin{equation}\label{obliv_start}
U\ket{0}^{\otimes m}\ket{\psi}=\sqrt{\lambda}\ket{0}^{\otimes m}V\ket{\psi}+\sqrt{1-\lambda}\ket{\Phi^\perp}.
\end{equation}
The target state is $\ket{\Phi}=\ket{0}^{\otimes m}V\ket{\psi}$. We note that for oblivious amplitude amplification to work, it is required of $U$ that $\lambda$ in Eq.~\eqref{obliv_start} be independent of $\ket{\psi}$. Given this setup, the question is, how many uses of $U$ do we need to make $\ket{\Phi}$ with probability $1-\delta^2$?

The naive classical attempt would measure the first $m$-qubits, succeeding with probability $\lambda$ in getting $\ket{\Phi}$. However, upon failure the state will generally be destroyed, and given only one copy of $\ket{\psi}$, we will not be able to retry except perhaps in special cases when $\ket{\psi}$ can be recovered from the remains after measurement. 

In \cite{berry2014exponential} Berry et~al.~give an exact quantum algorithm for oblivious amplitude amplification in the case of known $\lambda$ using the expected number $\mathcal{O}(1/\sqrt{\lambda})$ of uses of $U$. They argue that, if $\Pi=(\ket{0}\bra{0})^{\otimes m}\otimes I$, the operations $Ue^{i\pi\Pi}U^\dag$ and $e^{i\pi\Pi}$ act as the reflection about the start and reflection about the target states, respectively. Thus, Grover iterates constructed from these reflections act as expected on the 2-dimensional span of $\ket{\Phi}$ and $\ket{\Phi^\perp}$.

However, it is not immediately clear how to extend oblivious amplitude amplification to when $\lambda$ is unknown. Standard Grover algorithms in the unknown-$\lambda$ setting use a set of sequences of Grover iterates that increase exponentially in length, with a measurement after each to determine whether the target state has been found \cite{brassard2002quantum}. Each such measurement (now performed on the $m$-qubit ancilla) will, like in the classical approach, destroy the state if it fails. Moreover, at least one measurement is likely to fail during the process of ramping up the sequence lengths.

A workable approach is to conduct a coherent version of this exponential step-up in sequence length. For the problem of standard Grover search, this is essentially the idea from the last section of \cite{mosca2001counting}, where amplitude estimation is used. With great enough precision (a lower bound $w\le\lambda$ is required here to ensure enough precision), the estimation will, with high probability, yield a state that has overlap $1/2$ with the target state. Exact quantum search can from there complete the algorithm. Applying this approach to oblivious amplitude amplification, we get an algorithm that succeeds with probability $1-\delta^2$ using $\bigoh{1/(\delta\sqrt{w})}$ applications of $U$. Yet, not only is the scaling with $\delta$ poor, but there is also not a clear adiabatic version of this approach.

However, fixed-point search can also solve the problem of oblivious amplitude amplification with unknown $\lambda$ (and a known lower bound $w\le\lambda$). If the optimal fixed-point algorithm \cite{yoder2014fixed} is used this will take just $\bigoh{\log(1/\delta)/\sqrt{w}}$ uses of $U$. To see that this works is to realize that $Ue^{-i\alpha\Pi}U^\dag$ and $e^{i\beta\Pi}$ implement partial reflections about the start and target states, respectively, following exactly the same reasoning as in \cite{berry2014exponential}. Since fixed-point search in the gate-model is just unitary application of these partial reflections, it will produce the same amplification profile as a function of $\lambda$ as it does in the non-oblivious case. As an aside, a procedure using the digital-filter engineering that is also behind the fixed-point search \cite{yoder2014fixed} has recently improved the Hamiltonian simulation from \cite{berry2014exponential} by bypassing oblivious amplitude amplification entirely \cite{low2016optimal}.

An adiabatic oblivious search algorithm is also just as straightforward as the non-oblivious version, except with a modified Hamiltonian. One should construct $H=(1-s(t))H_B+s(t)H_E$, where now $H_E=\Pi$ and $H_B=U\Pi U^\dag$ (so that $e^{-iH_B}=Ue^{-i\Pi}U^\dag$). Assuming one can do this, the adiabatic search schedules $s(t)$ we have presented can be applied to the oblivious case without change. The modified $H_B$ may even be easier to implement than a full reflection about the initial state as you would find in a non-oblivious adiabatic search, as it can limit coupling between the $n$-qubit and $m$-qubit subsystems to just the bare essential coupling represented by $U$.

\section{conclusion}

Inspired by the fixed-point algorithm in \cite{yoder2014fixed}, we wondered whether the search algorithm in the adiabatic model could be made to have the fixed-point property. We found that the landscape of search algorithms in the adiabatic model is similar to that in the circuit model: some algorithms are fixed point but lack Grover-like scaling, some have Grover-like scaling but are not fixed point, and some have both properties. We have given examples for each of these cases. Moreover, we showed that simulation of a fixed-point adiabatic algorithm with Grover-like scaling can yield a fixed-point gate-model algorithm which retains the quantum speedup and acts as an alternative to the algorithms in \cite{aaronson2012quantum,yoder2014fixed}.

Fixed-point search algorithms provide fault-tolerance to certain systematic errors and misestimations of $\lambda$, and also find applications in certain problems with natural lower bounds, like preparation of the relatively-prime state, and even in counterfeiting quantum money \cite{aaronson2012quantum}. 

That fixed-point Grover-like algorithms exist in the adiabatic model extends these benefits to AQC but also gives us continuous control over the parameters of the algorithm $w$ and $\epsilon$ which we did not have in the gate-model. This continuous control might be useful to create a scheme which estimates the value of $\lambda$ by binary searching the interval $[0,1]$ using the adiabatic search algorithm with different parameters, or different schedule families altogether. Such a scheme would provide a novel way to perform quantum counting \cite{brassard1998quantum}, but also, potentially, a practical method for calibrating adiabatic quantum computers.

Direct applications aside, the method by which we proved our results is an equally important takeaway from this work. The technique was motivated by the intuition that a two-dimensional system exposed to a Hamiltonian is equivalent to a spin immersed in a magnetic field and can be treated geometrically on the Bloch sphere.
In this picture, the state vector precesses around the Hamiltonian vector, and by making a series of time-dependent coordinate changes we demonstrated an upper bound on the angle between the state and Hamiltonian vectors at the end of the algorithm.
For the fast schedule when $\lambda = w$, this method actually allowed for the Schr{\"o}dinger evolution to be solved exactly.

This approach contrasts with other rigorous methods of dealing with the adiabatic approximation, such as the asymptotic expansion of the error. Our bound is explicit and useful even in situations where the parameters are finite, giving us more than just asymptotic information about the algorithm. The ability to compute the error exactly (Theorem \ref{thm:fastexact}), albeit in a special case, is a unique result for AQC.

Further work might explore other families of schedules to achieve better dependence of $T$ on the error amplitude $\delta$, modify our geometric method to also find a lower bound on the error in the adiabatic approximation, or extend our ideas to bound the error in the adiabatic approximation in more than two dimensions.

\begin{acknowledgments}
We thank Guang Hao Low for interesting discussions and perspective especially concerning Hamiltonian simulation. A.M.D.~acknowledges the Paul E.~Gray UROP fund. T.J.Y.~thanks the NDSEG fellowship program. The authors also acknowledge funding by the NSF CUA, and NSF RQCC Project No.~1111337.
\end{acknowledgments}

\bibliographystyle{ieeetr}
\bibliography{paperbib}{}

\appendix


\section{Proof of Theorem \ref{thm:standardfixedpoint}} \label{sec:pfthmstandardfixedpoint}

Theorem \ref{thm:standardfixedpoint} follows from application of Theorem \ref{thm:generalbound} to the standard schedule $s_s(t;\epsilon_s,w)$ given by Eq. (\ref{eq:s}). 

Since $\dot{s}_s(t) = \epsilon_s \Delta_w^2$, and $\Delta_w(0)=1$, we evaluate $d_0$ and $d_1$ from Theorem \ref{thm:generalbound} as
\begin{eqnarray} 
d_0 &=& 2\epsilon_s \sqrt{\lambda(1-\lambda)} \\
d_1 &=& \int_0^{T_s} dt \left\lvert \frac{d}{dt}\left(\frac{\epsilon_s \Delta_w^2\sqrt{\lambda(1-\lambda)}}{\Delta_\lambda^3}\right)\right\rvert\\
&=&\int_0^1 ds_s \left\lvert \frac{d}{ds_s}\left(\frac{\epsilon_s \Delta_w^2\sqrt{\lambda(1-\lambda)}}{\Delta_\lambda^3}\right)\right\rvert
\end{eqnarray}
In the integral for $d_1$, we note that if not for the absolute value, we would be integrating a derivative and the answer would simply be the net change in the value of $c(s_s) := \epsilon_s \Delta_w^2\sqrt{\lambda(1-\lambda)}/\Delta_\lambda^3$ from $s_s=0$ to $s_s=1$ (note that $2c=\tan(\chi)$ with $\chi$ from (\ref{eq:chi})). The absolute value bars mean that we must count both positive and negative changes as positive, so we can evaluate the integral by observing intervals of $s_s$ over which $c(s_s)$ is monotonic, and summing the absolute value of the change of $c(s_s)$ over each of these regions. Now we analyze over what regions this $c(s_s)$ is monotonic. 

First, let
\begin{eqnarray}
c_0 &=& \epsilon_s \sqrt{\lambda(1-\lambda)}
\label{eq:chi0} \\
c_{1/2} &=& \frac{\epsilon_s w \sqrt{1-\lambda}}{\lambda}
\label{eq:chi12} \\
c_{\text{crit}} &=& \frac{2\epsilon_s \sqrt{\lambda}(1-w)^{3/2}}{3\sqrt{3(1-\lambda)(\lambda-w)}}
\label{eq:chic} \\
s_{\text{crit}} &=& \frac{1}{2} - \frac{1}{2}\sqrt{\frac{2\lambda-3w+\lambda w}{(1-\lambda)(1-w)}}
\end{eqnarray}
Note that $d_0 = 2c_0$. There are 3 cases.
\begin{enumerate}[(1)]
\item \emph{Case I}: $\lambda \leq 3 w/ (2+w)$. There is one local extremum: $c(s_s)$ increases from $c_0$ at $s_s=0$ to $c_{1/2}$ at $s_s=1/2$ and back to $c_0$ at $s_s=1$. 
\centerline{$d_0+d_1 = 2c_{1/2}$}
\item \emph{Case II}: $3 w/ (2+w) < \lambda < (1+2w)/3$. There are three local extrema: $c(s_s)$ increases from $c_0$ at $s_s=0$ to $c_\text{crit}$ at $s_s=s_\text{crit}$, decreases to $c_{1/2}$ at $s_s=1/2$, then increases back to $c_\text{crit}$ at $s_s=1-s_\text{crit}$ and finally decreases to $c_0$ at $s_s=1$.
\centerline{$d_0+d_1 = 4c_\text{crit} - 2c_{1/2}$}
\item \emph{Case III}: $\lambda \geq (1+2w)/3$. There is one local extremum:  $c(s_s)$ decreases from $c_0$ to $c_{1/2}$ at $s_s=1/2$ then increases back to $c_0$ at $s_s=1$.
\centerline{$d_0+d_1 = 4c_0 - 2c_{1/2}$}
\end{enumerate}
Thus we have explicitly calculated $d_0+d_1$ as a piecewise function of $\lambda$ (supposing constant $w$ and $\epsilon_s$). We would like to show that in each of the three cases, we have $d_0+d_1 \leq 2 \epsilon_s$.  

We can see this immediately in Case I, because  $\lambda \geq w$ implies $c_{1/2} \leq \epsilon_s$ so $d_0+d_1 \leq  2\epsilon_s$. Similarly for Case III, since $c_0 \leq \epsilon_s/2$ for all $\lambda$ we know that $d_0+d_1 \leq 4 c_0 \leq 2 \epsilon_s$.  Case II is difficult to analyze analytically, but the fact that $d_0+d_1 \leq 2 \epsilon_s$ for all $w$ and all $\lambda \in [3w(2+w),(1+2w)/3]$ can be verified graphically. This proves the theorem. \QEDA


\section{Proof of Theorem \ref{thm:simulatedgeneralbound}}

Since the maximum error amplitude $\delta$ is at most $1$, the theorem is trivially true when $\delta t \geq 2 \pi$. Thus we assume $\delta t < 2 \pi$ and treat $\delta t$ as a fixed input parameter. We define $H_t(s)$ as the solution to the equation
\begin{equation}
\exp(-i(1-s)H_0 \delta t)\exp(-i s H_1 \delta t) = \exp(-i H_t \delta t)
\label{eq:HtfromUt}
\end{equation}
To solve this equation, first we assume that $H_t$ has the form
\begin{equation}\label{eq:Htform}
H_t = \frac{1}{2}I-\frac{\gamma}{2}\hat{n}_t \cdot \vec{\sigma}
\end{equation}
where $\hat n_t$ is a unit vector.

Then, we write $H_0$ and $H_1$ using Eqs.~(\ref{eq:H0geo}) and (\ref{eq:H1geo}), expand both sides as a linear combination of $\{I,X,Y,Z\}$ using $\exp(i\theta \hat u \cdot \vec{\sigma})=\cos(\theta)I+i\sin(\theta)\hat u \cdot \vec{\sigma}$ (where $\hat u$ is a unit vector) and equate coefficients, arriving at
\begin{eqnarray}
\gamma &=& \frac{2}{\delta t} \arccos\left(\cos(a-b)-2\lambda\sin(a)\sin(b)\right) \label{eq:gamma} \\
\hat n_t \cdot \hat x &=& {\sin(\frac{\gamma \delta t}{2})}^{-1}\left(\sqrt{\lambda}\; \sin(a+b)\right)\label{eq:ntx} \\
\hat n_t \cdot \hat y &=& \sin(\frac{\gamma \delta t}{2})^{-1}\left(-2\sqrt{\lambda(1-\lambda)}\sin(a)\sin(b)\right) \label{eq:nty} \\
\hat n_t \cdot \hat z &=& \sin(\frac{\gamma \delta t}{2})^{-1} \left(\sqrt{1-\lambda}\sin(a-b)\right) \label{eq:ntz}
\end{eqnarray}
where $a=(1-s)\delta t/2$ and $b=s \delta t/2$. 

As expected, in the limit as $\delta t$ approaches 0, we can see that $\hat n_t$ approaches $\hat n$ and $\gamma$ approaches $\Delta_\lambda$.  For finite $\delta t$, $\hat n$ and $\hat n_t$ take approximately identical though slightly different paths. In particular, $\hat n$ stays in the $xz$-plane for all $s$, while $\hat n_t$ has a small component in the $\hat y$ direction.

We also discretize $H_t$ to form
\begin{equation}
H_{td}(t) = H_t(\lfloor t/\delta t \rfloor \delta t),
\label{eq:Htd}
\end{equation}
and we can see that the unitary evolution operator corresponding to application of $H_{td}$ between times $t_j$ and $t_{j+1}$ is precisely $U_t^{(j)}$ from Eq.~(\ref{eq:Utexp}).

Furthermore, we let $\ket{\psi_t(t)}$ and $\ket{\psi_{td}(t)}$ be states which evolve by Hamiltonians $H_t$ and $H_{td}$, respectively. So, $\ket{\psi_{td}(T)} = \ket{\psi_l}$, the final state of the simulated adiabatic search algorithm in the gate model, and hence $P(\lambda) = \lvert \braket{\psi_{td}(T)}{E} \rvert^2$. We bound this probability by first showing that $\lvert \braket{\psi_{t}(T)}{E} \rvert^2$ is close to 1, and then bounding the distance between $\ket{\psi_{t}(T)}$ and $\ket{\psi_{td}(T)}$.

\vspace{12 pt}


\noindent\textbf{Claim 1.} $\lvert \braket{\psi_{t}(T)}{E} \rvert^2 \geq 1-(\sqrt{2.6\delta t}+(d_0+d_1)(1+\delta t^2/25))^2$ if $\lambda \geq w$. 

\vspace{12 pt}


\noindent\textbf{Claim 2.} $\lvert \braket{\psi_{td}(T)}{\psi_t(T)} \rvert^2 \geq 1-2\delta t$


\begin{lemma}\label{lem:triangleinequality} 
If $\lvert \braket{Q}{S} \rvert^2 \geq 1-e_1^2$, and $\lvert \braket{R}{S} \rvert^2 \geq 1-e_2^2$, then $\lvert \braket{Q}{R} \rvert^2 \geq 1-(e_1+e_2)^2$ for any normalized $\ket Q$, $\ket R$, $\ket S$.
\end{lemma}


We delay the proof of Lemma \ref{lem:triangleinequality} until the end of the section. Since $\sqrt{2}+\sqrt{2.6} < 3.1$, Claims 1 and 2 combined with Lemma \ref{lem:triangleinequality} are sufficient to prove Theorem \ref{thm:simulatedgeneralbound}. Now we explain why each claim is true.

\vspace{12 pt}


\begin{mainproof}{Proof of Claim 1} Claim 1 is an upper bound on the error resulting from running the continuous search algorithm with Hamiltonian $H_t$ instead of $H$. The claim highly resembles Theorem \ref{thm:generalbound}, and a very similar technique is used in the following proof. 

As in Theorem \ref{thm:generalbound}, we make a time-dependent change of coordinates via a time-dependent unitary transformation so that the Hamiltonian vector is rotated back to the $\hat z$ axis. In this case, $\hat n_t$ has components in all 3 directions, so constructing this unitary is more complicated. We express this transformation as a product of 3 unitaries. 
\begin{equation}
\ket {\phi_t(t)} = U_3(t)U_2(t)U_1(t) \ket{\psi_t(t)}
\label{eq:ketphit}
\end{equation}
with $U_1 = \exp(i \frac{\eta_1}{2} \hat n \cdot \vec{\sigma})$, 
$U_2 = \exp(-i \frac{\eta_2}{2} (\hat n \times \hat y )\cdot \vec{\sigma})$, and $U_3 = \exp(i\frac{\theta}{2} Y)$. $U_1$ is a rotation about $\hat n$ by an angle $\eta_1$, $U_2$ is a rotation about the vector perpendicular to $\hat n$ in the xz-plane by an angle of $\eta_2$. We will choose $\eta_1$ and $\eta_2$ so these two operations alone rotate the vector $\hat{n}_t$ onto the vector $\hat{n}$. Then, $U_3$ is the same unitary we used to prove Theorem \ref{thm:generalbound} that rotates $\hat n$ back to $\hat z$. We find the rotation angles we need are 
\begin{eqnarray}
\eta_1 &=& \arctan\left(\frac{(\hat n_t \cdot \hat x)\cos(\theta)-(\hat n_t \cdot \hat z)\sin(\theta)}{\hat n_t \cdot \hat y} \right) \label{eq:eta1} \\
\eta_2 &=& \arccos\left((\hat n_t \cdot \hat x)\sin(\theta)+ (\hat n_t \cdot \hat z)\cos(\theta)\right) \label{eq:eta2}
\end{eqnarray}
where the components of $\hat n_t$ are given by eqs: (\ref{eq:ntx}-\ref{eq:ntz}). Recall from Eq.~(\ref{eq:theta}) that $\cos(\theta) = (1-2s)\sqrt{1-\lambda}/\Delta_\lambda = \hat n \cdot \hat z$ and $\sin(\theta) = \sqrt{\lambda}/\Delta_\lambda = \hat n \cdot \hat x$. Lemma \ref{lem:unitaryrotation} verifies that this transformation works.


\begin{lemma}\label{lem:unitaryrotation}
\begin{equation}
(U_3U_2U_1)H_t(U_3U_2U_1)^\dagger = \frac{I}{2}-\frac{\gamma}{2}Z
\label{eq:conjugationHt}
\end{equation}
\end{lemma}
%
%

Moreover, when $s=1$ we have $\gamma = 1$, $\ket{\hat n} = \ket E$, and $\eta_2=0$, so up to a global phase $U_2U_1\ket E = \ket E$ and $U_3U_2U_1\ket E = \ket 0$. Hence, we can rewrite $\lvert \braket{\psi_{t}(T)}{E} \rvert^2= \lvert \braket{0}{\phi_t(T)}\rvert^2$. Now we calculate the effective Hamiltonian by which the state $\ket {\phi_t(t)}$ evolves. Note that $\lVert O \rVert$ denotes the maximum norm of $O\ket \alpha$ over all normalized states $\ket \alpha$.


\begin{lemma}\label{lem:schrodingerphit}
$\ket {\phi_t(t)}$ evolves according to $i \frac{d}{dt} \ket {\phi_t(t)} = H_{\phi t} \ket{\phi_t(t)}$
where
\begin{equation}
H_{\phi t} = \frac{I}{2} - \frac{\gamma}{2}Z - \frac{\dot \theta}{2} Y + E 
\label{eq:Hphit}
\end{equation}
and $\int_0^T \lVert E \rVert dt \leq 1.3\delta t$.
\end{lemma}


Since $\gamma$ is approximately $\Delta_\lambda$, $H_{\phi t}$ is approximately $H_\phi$ from Eq.~(\ref{eq:Hphi}) plus a bounded error term $E$. We write $H_{\phi t} = H_{\phi t 0} + E$. If $\ket {\phi_{t0}}$ evolves according to $H_{\phi t 0}$, we wish to quantify how far away $\ket {\phi_{t0}}$ is from $\ket {\phi_{t}}$ at time $T$, which is facilitated by Lemma \ref{lem:perturbation}.


\begin{lemma}\label{lem:perturbation} If $\ket {\Psi_A(t)}$ evolves by Hamiltonian $H_A$, while $\ket {\Psi_B(t)}$ evolves by Hamiltonian $H_B$, and $\ket {\Psi_A(0)}=\ket {\Psi_B(0)}$ then $\lvert \braket{\Psi_A(T)}{\Psi_B(T)} \rvert \geq 1-\int_0^T \lVert H_A-H_B \rVert dt$.
\end{lemma}


Lemmas \ref{lem:schrodingerphit} and \ref{lem:perturbation} tell us that 
\begin{equation}
\lvert \braket{\phi_{t0}(T)}{\phi_{t}(T)} \rvert^2 \geq 1-2.6 \delta t
\label{eq:roterror}
\end{equation}
%

\begin{lemma}\label{lem:trotgeneralbound} $\lvert \braket{\phi_{t0}(T)}{0} \rvert^2 \geq 1-((d_0+d_1)(1+ \delta t^2/25))^2$ if $\lambda \geq w$, where $d_0$ and $d_1$ are given by Eqs.~(\ref{eq:generalboundd0}) and (\ref{eq:generalboundd1}) in Theorem \ref{thm:generalbound}. 
\end{lemma}


Compare this result to the conclusion of Theorem \ref{thm:generalbound}, that $\lvert \braket{\phi(T)}{0} \rvert^2 \geq 1-(d_0+d_1)^2$. We have now bounded in Eq.~(\ref{eq:roterror}) how close $\ket{\phi_{t0}(T)}$ and $\ket{\phi_{t}(T)}$ are, and in Lemma \ref{lem:trotgeneralbound}, how close $\ket{\phi_{t0}(T)}$ and $\ket 0$ are. We can again use Lemma \ref{lem:triangleinequality} to conclude that 
\begin{equation}
\lvert \braket{\phi_t(T)}{0} \rvert^2 \geq 1-(\sqrt{2.6\delta t}+(d_0+d_1)(1+\delta t^2/25))^2
\label{eq:continuousdelta}
\end{equation}
whenever $\lambda \geq w$. This proves Claim 1. We have shown that if the state is exposed to $H_t$, instead of $H$, we still have a search algorithm with bounded error probability that is related only to $\delta t$ and the schedule $s(t)$. If we relax the fixed-point definition to allow algorithms whose error can be bounded by functions of both $\epsilon$ and $\delta t$ (but not $w$), then applying $H_t(s(t;\epsilon,w))$ would be a fixed-point algorithm for any schedule for which applying $H(s(t;\epsilon,w))$ is a fixed-point algorithm. Claim 2 illustrates how the additional error incurred by using the discrete Hamiltonian $H_{td}$ instead of $H_t$ can be bounded as well.
\end{mainproof}



\begin{mainproof}{Proof of Claim 2} 
First we examine the definition of $H_t$ to bound $\lVert dH_t/ds \rVert$. We differentiate both sides of Eq.~(\ref{eq:HtfromUt}) and cancel factors arriving at
\begin{eqnarray}  
\frac{dH_t}{ds} &=& - H_0 + e^{-i(1-s)H_0\delta t}H_1 e^{i(1-s)H_0\delta t} \nonumber \\
&=& -\hat n_0 \cdot \vec{\sigma}/2 +  e^{-i(1-s)H_0\delta t}(\hat n_1 \cdot \vec{\sigma}) e^{i(1-s)H_0\delta t}/2 \nonumber\\
\left\lVert \frac{dH_t}{ds}\right\rVert & \leq & \lVert\hat n_0 \cdot \vec{\sigma} \rVert/2 + \lVert \hat n_1 \cdot \vec{\sigma} \rVert/2 = 1
\end{eqnarray}

We define $\delta H_t = H_t-H_{td}$. $\delta H_t(t_j) = 0$ for all $j$, and if $t_j \leq t < t_{j+1}$, then
\begin{eqnarray} 
\lVert \delta H_t(t) \rVert = (t-t_j)\frac{d}{dt}\lVert \delta H_t(c) \rVert \leq \delta t \frac{d}{dt}\lVert \delta H_t(c) \rVert
\end{eqnarray} 
for some $c$ with $t_j < c < t$. But 
\begin{equation} 
\frac{d}{dt}\lVert \delta H_t(c) \rVert = \frac{ds}{dt}\frac{d}{ds}\lVert \delta H_t(c) \rVert \leq \frac{ds}{dt}\lVert\frac{d}{ds} \delta H_t(c) \rVert \leq \frac{ds}{dt}
\end{equation}
So, now we have
\begin{equation}
\int_0^T \lVert \delta H_t \rVert dt \leq \int_0^T \delta t\frac{ds}{dt} dt = \delta t
\end{equation}
Again invoking Lemma \ref{lem:perturbation}, we arrive immediately at the Claim.
\end{mainproof}



\begin{mainproof}{Proof of Lemma \ref{lem:triangleinequality}} Since all states are qubits, this follows from the triangle inequality over a sphere. If the angle between states $\ket Q$ and $\ket R$ is $2\nu_{qr}$, between $\ket Q$ and $\ket S$ is $2\nu_{qs}$, and between $\ket R$ and $\ket S$ is $2\nu_{rs}$, then we can say that $\lvert\braket{Q}{R}\rvert^2 = \cos^2(\nu_{qr})$, $\lvert\braket{Q}{S}\rvert^2 = \cos^2(\nu_{qs})$ and $\lvert\braket{R}{S}\rvert^2 = \cos^2(\nu_{rs})$. So $e_1 = \sin(\nu_{qs})$ and $e_2 = \sin(\nu_{rs})$. The triangle inequality says that $\nu_{ab} \leq \nu_{ac}+\nu_{bc}$, and therefore
\begin{eqnarray}
\sin(\nu_{qr}) &\leq & \sin(\nu_{qs}+\nu_{rs}) \nonumber \\
&\leq & \sin(\nu_{qs})\cos(\nu_{rs}) + \sin(\nu_{rs})\cos(\nu_{qs}) \nonumber \\
&\leq & \sin(\nu_{qs}) + \sin(\nu_{rs}) \nonumber \\
& \leq & e_1 + e_2 \nonumber \\
\lvert \braket{Q}{R} \rvert^2 &=&1-\sin^2(\nu_{qr})\geq 1-(e_1+e_2)^2 \nonumber
\end{eqnarray}
and the Lemma is proved. \end{mainproof}



\begin{mainproof}{Proof of Lemma \ref{lem:unitaryrotation}} We will rely on the identity
\begin{equation}
\exp(-i\,\alpha\,\hat r \cdot \vec\sigma)( \vec v \cdot \vec\sigma)\exp(i\,\alpha\,\hat r \cdot \vec\sigma) = R_{\hat r}(2\alpha)\vec{v} \cdot \sigma
\label{eq:operatorrotation}
\end{equation}
where $R_{\hat r}(2\alpha)$ represents the operator which rotates vectors about $\hat r$ through an angle of $2\alpha$. $H_t$ undergoes three subsequent rotations. However, it will be helpful to rearrange the rotations as follows:
\begin{eqnarray}  
U_3U_2U_1 &=& (U_3U_2U_3^\dagger)(U_3U_1U_3^\dagger)U_3 \nonumber \\
U_3U_2U_3^\dagger&=&\exp\left(-i \frac{\eta_2}{2}U_3((\hat n \times \hat y) \cdot \vec\sigma)U_3^\dagger\right) \nonumber \\
&=&\exp\left(-i \frac{\eta_2}{2}X\right) \nonumber \\
U_3U_1U_3^\dagger &=&\exp\left(-i \frac{\eta_1}{2}U_3(\hat n \cdot \vec\sigma)U_3^\dagger\right) \nonumber \\
&=& \exp\left(-i \frac{\eta_1}{2}Z\right) \nonumber \\
\therefore U_3U_2U_1&=& \exp\left(-i \frac{\eta_2}{2}X\right)\exp\left(-i \frac{\eta_1}{2}Z\right)\exp\left(i\frac{\theta}{2}Y\right) \nonumber
\end{eqnarray}
So, expanded this way, the first rotation is about the $y$-axis. If the Hamiltonian starts in direction $\hat n_t$, it will finish in direction 
\begin{eqnarray}  
\hat r_t = ((\hat n_t \cdot \hat x) \cos(\theta) - (\hat n_t \cdot \hat z) \sin(\theta))&\hat x&  \nonumber \\
+\; (\hat n_t \cdot \hat y) &\hat y& \nonumber \\
 +\; ((\hat n_t \cdot \hat x) \sin(\theta) + (\hat n_t \cdot \hat z) \cos(\theta) )&\hat z&
\nonumber
\end{eqnarray}
Now, if we rotate $\hat r_t$ about the $z$-axis by an angle of $\eta_1 = \arctan((\hat r_t \cdot x)/(\hat r_t \cdot y))$, we will preserve the $z$-component of $\hat r_t$ while rotating the $xy$-part onto the $y$-axis. Then, we rotate about the $x$-axis an angle $\eta_1 = \arccos(\hat r_t \cdot \hat z)$ which will rotate it back to the $z$-axis. Since the norm is preserved by all of these rotations, the final result is $I/2 - \gamma Z/2$ as claimed.
\end{mainproof}



\begin{mainproof}{Proof of Lemma \ref{lem:schrodingerphit}} If we plug Eq.~(\ref{eq:ketphit}) 
into the Schr{\"o}dinger equation $id\ket{\psi_t}/dt = H_t \ket{\psi_t}$, we find that $\ket{\phi_t}$ evolves according to Hamiltonian
\begin{eqnarray} 
H_{\phi t} &=& i \frac{dU_3}{dt}U^\dagger_3+ i U_3\frac{dU_2}{dt}U^\dagger_2U^\dagger_3  + i U_3U_2\frac{dU_1}{dt}U^\dagger_1U^\dagger_2U^\dagger_3 \nonumber \\
&&+ U_3U_2U_1H_tU^\dagger_1U^\dagger_2U^\dagger_3 \nonumber 
\end{eqnarray}
The $Y$ term in Eq.~(\ref{eq:Hphit}) comes immediately from the first term above: 
\begin{equation} 
i\frac{dU_3}{dt}U^\dagger_3 = -\frac{\dot{\theta}}{2}Y \nonumber
\end{equation}
The $I$ and $Z$ terms will come from the conjugation of $H_t$ as shown in Lemma \ref{lem:unitaryrotation}. 
So we have 
\begin{eqnarray} 
E &=& i U_3\frac{dU_2}{dt}U^\dagger_2U^\dagger_3 + i U_3U_2\frac{dU_1}{dt}U^\dagger_1U^\dagger_2U^\dagger_3 \nonumber \\
\lVert E \rVert &\leq & \left\lVert \frac{dU_2}{dt} \right\rVert + \left\lVert \frac{dU_1}{dt} \right\rVert \nonumber
\end{eqnarray}
and $U_1 = \cos(\eta_1/2) I + i \sin(\eta_1/2)\hat{n}\cdot \vec{\sigma}$, so $dU_1/dt = -\dot{\eta}_1 \sin\left(\eta_1/2\right)I/2 + i\left(\dot{\eta}_1\cos(\eta_1/2\right)\hat{n}/2+\sin\left(\eta_1/2\right)\dot{\hat{n}})\cdot\vec{\sigma}$ whose norm is given by
\begin{eqnarray} 
&&\left\lVert \frac{dU_1}{dt} \right\rVert \nonumber \\ &=&\sqrt{\frac{\dot{\eta}_1^2}{4}\sin^2\left(\frac{\eta_1}{2}\right)+\frac{\dot{\eta}_1^2}{4}\cos^2\left(\frac{\eta_1}{2}\right)+\lvert \dot{\hat{n}} \rvert^2\sin^2\left(\frac{\eta_1}{2}\right) } \nonumber \\
&\leq& \left\lvert\frac{\dot{\eta}_1}{2}\right\rvert + \left\lvert \dot{\hat{n}}\sin\left(\frac{\eta_1}{2}\right)\right\rvert \leq \frac{1}{2}(\lvert\dot{\eta}_1\rvert + \lvert \eta_1\dot{\hat n} \rvert) \nonumber
\end{eqnarray}
Where the second line follows since $\hat n$ is orthogonal to $\dot{\hat n}$. Likewise,
\begin{equation} 
\left\lVert \frac{dU_2}{dt} \right\rVert \leq \frac{1}{2} \lvert \dot \eta_2 \rvert + \left\lvert \frac{\eta_2}{2}\frac{d}{dt}(\hat n \times \hat y)\right\rvert = \frac{1}{2}(\lvert \dot \eta_2 \rvert + \lvert\eta_2\dot{\hat n}\rvert) 
\end{equation}
with the final equality holding since $\hat n$, $\dot{\hat n}$, and $\hat y$ are mutually orthogonal. 

Graphically, we can verify (can be justified analytically), that $\eta_1$ is monotonically decreasing as a function of $s$ with $\eta_1(s=0) = \eta_{1,\max}$ and $\eta_1(s) = -\eta_1(1-s)$ for all $s$. Meanwhile, $\eta_2$ monotonically increases from $\eta_2(s=0)=0$ to its maximum $\eta_2(s=1/2)=\eta_{2,\max}$, and then returns to $\eta_2(s=1)=0$, maintaining the symmetry $\eta_2(s) = \eta_2(1-s)$ for all $s$.
\begin{eqnarray} 
&&\int_0^T \lVert E \rVert dt \nonumber \\
&\leq & \frac{1}{2}\int_0^T (\lvert \dot \eta_1 \rvert + \lvert \dot \eta_2 \rvert)dt + \frac{1}{2}\int_0^T\lvert \dot {\hat n}\rvert (\vert\eta_1\rvert+\lvert\eta_2\rvert) dt \nonumber \\
&\leq & \eta_{1,\max} +\eta_{2,\max} + \frac{\eta_{1,\max}+\eta_{2,\max}}{2}\int_0^T \lvert \dot{\hat n} \rvert dt \nonumber \\
& \leq & (1+\pi/2)(\eta_{1,\max}+\eta_{2,\max})
\end{eqnarray}
since $\int_0^T \lvert \dot{\hat n} \rvert dt$ is the arclength of the path traversed by $\hat n$, which is always less than $\pi$. Now, we can verify that $\eta_{1,\max} = \eta_1(s=0) \leq \delta t/4$ and $\eta_{2,\max} = \eta_2(s=1/2) \leq \delta t/4$ as long as $\delta t \leq 2 \pi$ (analytically or graphically). Thus, $\int_0^T \lVert E \rVert dt \leq (1+\pi/2)(\delta t/2) \leq 1.3 \delta t$, which proves the Lemma.
\end{mainproof}



\begin{mainproof}{Proof of Lemma \ref{lem:perturbation}} We let $C = \braket{\Psi_0(T)}{\Psi_1(T)} $, and we differentiate both sides
\begin{eqnarray} 
\frac{dC}{dt} &=& -i\bra {\Psi_0(T)}H_1 \ket {\Psi_1(T)}+i\bra {\Psi_0(T)}H_0 \ket {\Psi_1(T)} \nonumber \\
\left\lvert \frac{dC}{dt} \right\rvert &=& \lvert \bra {\Psi_0(T)}H_1-H_0 \ket {\Psi_1(T)}\rvert \leq \lVert H_1-H_0 \rVert \nonumber
\end{eqnarray}
and the lemma follows immediately.
\end{mainproof}



\begin{mainproof}{Proof of Lemma \ref{lem:trotgeneralbound}} We go through the same process used to prove Theorem \ref{thm:generalbound}, except most of the work has already been done for us. We define 
\begin{equation}
\chi_t(t) = \arctan\left(\frac{ \dot{\theta}(t)}{\gamma(t)}  \right)
\label{eq:chit} 
\end{equation}
to mirror Eq.~(\ref{eq:chi}). And, as in Theorem \ref{thm:generalbound}, we can make the following bound
\begin{eqnarray} 
\lvert \braket{0}{\phi_{t0}} \rvert^2 &\geq & \cos^2(A_t) \geq 1-A_t^2 \nonumber \\
A_t &=& \chi_t(0) + \frac{1}{2} \int_0^T \lvert \dot \chi_t \rvert dt
\end{eqnarray}
Also, since $d(\arctan(u))/dt = \dot u/(1+u^2)\leq \dot u$, we can say that
\begin{equation} 
A_t \leq \frac{ \dot{\theta}(0)}{\gamma(0)} + \frac{1}{2} \int_0^T \left\lvert \frac{d}{dt}\left(\frac{ \dot{\theta}(t)}{\gamma(t)}\right) \right\rvert dt
\end{equation}
We not the following, where the subscript $\max$ denotes a quantities maximum value with $0\leq s \leq 1$. 
\begin{eqnarray} 
\left(\frac{ \dot{\theta}}{\gamma}\right) &=& \left(\frac{ \dot{\theta}}{\Delta_\lambda}\right) \left(\frac{ \Delta_\lambda}{\gamma}\right) \nonumber \\
\partial_t\left(\frac{ \dot{\theta}}{\gamma}\right)  &=& \partial_t\left(\frac{\dot{\theta}}{\Delta_\lambda}\right) \left(\frac{ \Delta_\lambda}{\gamma}\right)+\left(\frac{\dot{\theta}}{\Delta_\lambda}\right) \partial_t\left(\frac{ \Delta_\lambda}{\gamma}\right) \nonumber \\
\left\lvert\partial_t\left(\frac{ \dot{\theta}}{\gamma}\right)\right\rvert  & \leq & \left\lvert\partial_t\left(\frac{\dot{\theta}}{\Delta_\lambda}\right) \right\rvert\left(\frac{ \Delta_\lambda}{\gamma}\right)_{\max} \nonumber\\
&&+\left(\frac{\dot{\theta}}{\Delta_\lambda}\right)_{\max} \left\lvert\partial_t\left(\frac{ \Delta_\lambda}{\gamma}\right)\right\rvert \nonumber
\end{eqnarray}
which allows us to rewrite
\begin{eqnarray} 
A_t &\leq & \left(\frac{ \dot{\theta}(0)}{\Delta_\lambda(0)} + \frac{1}{2} \int_0^T \left\lvert \partial_t\left(\frac{ \dot{\theta}(t)}{\Delta_\lambda(t)}\right) \right\rvert dt\right)\left(\frac{ \Delta_\lambda}{\gamma}\right)_{\max}  \nonumber \\
&& + \frac{1}{2}\left(\frac{\dot{\theta}(t)}{\Delta_\lambda(t)}\right)_{\max} \int_0^T \left\lvert \frac{d}{dt}\left(\frac{\Delta_\lambda(t)}{\gamma(t)}\right) \right\rvert dt
\end{eqnarray}
If we look at $\Delta_\lambda(s)/\gamma(s)$, we can see that it is always at least 1, increasing to a maximum at $s=1/2$, then decreasing back to $1$. If $(\Delta_\lambda(s)/\gamma(s))_{\max} = 1+x$. Graphically, we can see that $x\leq \delta t^2/50$ whenever $\delta t \leq 2\pi$ (can be justified analytically). Moreover, Theorem \ref{thm:generalbound} tells us that the first term is at most $d_0+d_1$ given by Eqs.~(\ref{eq:generalboundd0}) and (\ref{eq:generalboundd1}). Also, for any quantity $u$, we have $u_{\max} \leq u(0)+\int \lvert u \rvert$, so  $(\dot{\theta}(t)/\Delta_\lambda(t))_{\max} \leq d_0+d_1$ as well. Therefore, we have
\begin{eqnarray} 
A_t & \leq & (d_0+d_1)(1+x)+ (d_0+d_1)x \nonumber \\
\lvert \braket{\phi_{t0}}{0} \rvert^2 &\geq& 1-((d_0+d_1)(1+2x))^2
\end{eqnarray}
Using $\delta t^2/50$ for $x$ gives the Lemma.
\end{mainproof}



\end{document}